\documentclass[twocolumn]{article}

\usepackage[utf8]{inputenc}
\usepackage[T1]{fontenc}
\usepackage{multicol}
\usepackage{lipsum} 
\pdfoutput=1 

\usepackage[T1]{fontenc} 

\usepackage{graphicx}
\usepackage{subcaption}
\usepackage{amsmath} 
\usepackage{csquotes} 
\usepackage{verbatim} 
\usepackage{mathrsfs}
\usepackage{amssymb}
\usepackage[dvipsnames]{xcolor}
\usepackage{amsfonts}
\usepackage{graphics}
\usepackage{epsfig}
\usepackage{hyperref}
\usepackage{multicol}
\usepackage{caption}
\newcommand{\bra}{\langle}
\newcommand{\ket}{\rangle}
\newcommand{\mH}{\mathcal{H}}
\newcommand{\nk}{\textbf{k}}

\newcommand{\dphi}{\delta \phi}
\newcommand{\mR}{\mathcal{R}}
\newcommand{\nq}{\textbf{q}}
\newcommand{\mP}{\mathcal{P}}
\newcommand{\x}{\textbf{x}}

\newcommand{\nn}{\nonumber \\}
\newcommand{\RI}{\text{R,I}}

\begin{document}

\twocolumn[
\begin{center}
{\LARGE Breaking Eternal Inflation: Empirical Viability of a Spontaneous Collapse Scenario \par}
\vspace{0.5em}

{\large Mar\'{\i}a P\'{\i}a Piccirilli$^{a,b}$, Gabriel Le\'{o}n$^{a,b}$, Rosa-Laura Lechuga-Solis$^{c,d,e}$, Daniel Sudarsky$^{e,f}$ \par}
\vspace{0.5em}

$^a$ Grupo de Cosmolog\'{\i}a, Facultad de Ciencias Astron\'{o}micas y Geof\'{\i}sicas, Universidad Nacional de La Plata, Argentina \\ 
$^b$ CONICET, Ciudad Aut\'{o}noma de Buenos Aires, Argentina \\ 
$^c$ Dipartimento di Fisica Galileo Galilei, Università di Padova, Italy \\ 
$^d$ INFN Sezione di Padova, Italy \\ 
$^e$ Instituto de Ciencias Nucleares, Universidad Nacional Autónoma de México, Mexico City, Mexico \\ 
$^f$ Departament de Física Quantica i Astrofísica, Universidad de Barcelona, España \\
\vspace{0.5em}

\texttt{mpp@fcaglp.unlp.edu.ar, gleon@fcaglp.unlp.edu.ar, rosalaura.lechugasolis@unipd.it, sudarsky@correo.nucleares.unam.mx}
\end{center}

\begin{abstract}
We revisit an inflationary scenario in which primordial inhomogeneities arise from a quantum collapse, a stochastic mechanism described in the context of quantum collapse theories in its continuous version and within semiclassical gravity. The predictions of the model \cite{Canate:2012nwv, RL2023} show a non-conventional scalar spectrum governed by two new parameters in the collapse rate, whose aim is twofold: on one side, to account for the primordial cosmic structure, and on the other to explain the suppression amplitude associated with long-wavelength modes, thereby eliminating the occurrence of eternal inflation. Furthermore,  this model can contribute to accounting for the lack of power anomaly in the low  $l$  angular power spectra of the Cosmic Microwave Background (CMB). Using the latest data from the \textit{Planck} (2018) collaboration, we establish observational constraints on the model parameters, which produce a characteristic low-$\ell$ suppression in the cosmic microwave background spectrum. We conclude that the \textit{Planck data} support the solution presented in \cite{RL2023}, in other words, that the model allows us to solve simultaneously the emergence of the cosmic structure and, at the same time, avoid the eternal inflation scenario.
\end{abstract}

\vspace{2em} 
] 

\section{Introduction}
The second, and latest, data release (DR2) BAO results reported by DESI are broadly consistent with the so-called $\mathbf{{\Lambda}{{\textbf{CDM}}}}$ model \cite{DESI:2025zpo}. Such a model includes the inflationary mechanism, a period of accelerated expansion in the early universe, which was supported by Planck data in 2018 \cite{Planck2018Results}. As is well known, beyond solving the naturalness problems \cite{Guth:1980zm}, inflation is supposed to account for the emergence of the seeds of primordial cosmic structure. However, the question regarding the explanation of the breaking of symmetry (homogeneity and isotropy) of the Bunch-Davies vacuum remains open. One approach that has been explored to address this issue incorporates spontaneous quantum collapse as a stochastic mechanism which can, in fact, account for the symmetry breaking \cite{pss}. The incorporation of this aspect involves a modification of standard quantum mechanics, as the cosmological setting is one where the inclusion of the influence of external observers as justification for the reliance on the reduction mechanism and departure from a standard unitary evolution is simply unviable. Spontaneous collapse theories, particularly the Continuous Spontaneous Localization version, CSL, are designed as general schemes to address the broader quantum mechanical measurement problem \cite{Norsen2017}, and have been previously considered in the cosmological context to deal with this conceptual problem we have just discussed. As shown in \cite{Canate:2012nwv}, it is possible to obtain a transparent and satisfactory conceptual explanation for the emergence of the primordial seeds structure, including the reproduction of the scale-invariant primordial Power Spectrum. A second study, \cite{piarunning}, carried out the same analysis up to second order in the slow-roll parameters, and such a Spectrum was also adequately reproduced and matched with observations \cite{piarunning2}.

A different, but no less important, caveat that arises in the standard treatment of the inflationary universe concerns the issue of dealing with a theoretical context involving both gravity and quantum field theory, given that we lack a sufficiently developed and generally usable theory of quantum gravity. It is customary to consider the inflationary period as dominated by a scalar field $\phi$, known as the inflaton, which is expressed as $\phi=\phi_{0}(\eta)+\delta\phi(\x,\eta)$. The background $\phi=\phi_{0}(\eta)$ is spatially constant and the spatial dependence is encoded in the perturbation $ \delta\phi(\x,\eta)$. For the space-time metric, one writes $g_{ab}=g_{ab}^{0}+\delta g_{ab}$, with the background $g_{ab}^{0}$ corresponding to the FLRW metric, and the perturbation encodes the geometric aspects of the structure. In those treatments the background inflaton field $\phi_{0}$, more precisely the corresponding potential energy density, drives the universe into the accelerated expansion regime, while the parts containing the explicit spatial dependencies $\delta\phi$, $\delta g_{ab}$ are treated in a quantum mechanical manner, in which, as a natural assumption for the initial state for the quantum modes, one takes that so-called Bunch-Davies vacuum. These modes are taken to describe the emergence of the primordial cosmic structure, as well as the primordial gravity waves.

In order to represent the inhomogeneities, the usual approaches relies on the study of the two-point function of quantum variables related to both the gravitational and matter degrees of freedom — in particular the Mukhanov-Sasaki variable, $v=a\left(\Psi+\frac{\mathcal{H}}{\phi_{0}^{'}}\delta \phi\right)$— which plays a central role in the characterization of the cosmic structure.

The treatment underlying these calculations is problematic for various reasons: \textbf{i)} the treatment of the metric perturbations relies on the methods of quantum field theory in curved space-time, and \textbf{ii)} the interpretation of quantum uncertainties of two-point functions as if they represented stochastic fluctuations of the underlying variables.

According to quantum field theory, the causality conditions constrain the form of the commutation relations of the quantum field operators, and, as stated in \cite[p. 384]{Wald:1984rg}, when one splits the metric into background and perturbation parts, one faces the issue of which is the space-time defining the causal structure to be used in that construction. 
On the one hand, one could simply rely on the causal structure of the background metric, which is well defined, as that sector is not subject to a quantum treatment. However, this does not seem like a reasonable path, if one really wants to take the space-time structure to be that associated with the ``full metric''. On the other hand, one might be inclined to rely on the ``full space-time metric,'' but then would face the problem that it is a quantum object and thus for general states no clear structure can be assigned. 
As mentioned before, in usual treatments of early-universe cosmology, it is customary to quantize both the matter and the space-time perturbations without much attention to the issue, and relying in practice on the first alternative. However, if we aim to obtain a conceptually clear explanatory account, it would not be ``legitimate'' just to ignore this aspect.

The former problem is intimately related to a broader issue affecting quantum theory in general, and known as the measurement problem. A key feature of quantum theory is that it involves two distinct laws of evolution: the unitary Schrödinger evolution of the quantum state, taken to apply when the system is left to evolve autonomously, and the reduction postulate, which governs the state’s change when a measurement is performed. One of the central aspects of the problem in general is that the theory is unclear regarding what counts as a measurement \cite{Bell}. Indeed, in the early cosmological context at hand, the issue becomes both simpler and at the same time more acute given the impossibility of identifying any entity that could reasonably play the role of  ``observer'' in epochs previous to the formation of galaxies, stars, and planets. Along with this, the fact that, in any event, one of the aspects one wants to account for, in this setting, is the very formation of the corresponding seeds of cosmic structure (these issues have been discussed at length \cite{Perez:2005gh} and the interested reader is referred to those works) that make life and real observers possible to start with. The short outcome of such studies is that, in the context of standard quantum theory is simply illegitimate to regard quantum uncertainties as if they were classical stochastic outcomes lacking anything that can be considered a measurement. In particular we should emphasize that the Bunch-Davies or ``adiabatic vacuum'' is completely homogeneous and isotropic quantum state (on the QFT associated with the equally homogeneous and isotropic background), and, as the standard quantum dynamics preserves these symmetries it is illegitimate to read into that quantum state (taken, as is usual in quantum theory, to provide a complete characterization of the system) any feature indicating anisotropies or inhomogeneities \cite{Sudarsky:2009za, Berjon:2020vdv}.


One framework that might be used to describe the emergence of the primordial structure is semi-classical gravity, in which matter fields are treated according to quantum theory while gravity is described at the classical level. As can already be seen from works such as \cite{Page}, this framework must be modified in a suitable manner in order to incorporate spontaneous collapses of the quantum states, just to avoid clear conflicts with empirical evidence. Such modifications are rather nontrivial (see the works cited below), as collapses are usually tied to violations of the divergence-free condition for the expectation value of the energy-momentum tensor (as already noted in \cite{Page} and \cite{Eppley1977}).
It is important to highlight that we are working with such semi-classical gravity versions, considering them as effective theories, in analogy to the role that the Navier–Stokes equations play in hydrodynamics (where here, and in contrast with that case, we simply lack what would play the role of the more fundamental theory). The point is that although the Navier–Stokes equations are not fundamental, failing to capture the underlying atomic and molecular structure of fluids, they do nevertheless provide a remarkably accurate description of macroscopic fluid behavior. However, such effective descriptions break down once excitations of the microscopic degrees of freedom become significant enough to invalidate the hydrodynamic approximation. 

In \cite{Canate:2012nwv, piarunning}, the emergence of primordial cosmic structure is addressed by employing an approach first studied in some detail in \cite{Diez:2012}, which combines collapse theories, together with semi-classical gravity. The works \cite{Canate:2012nwv, piarunning} rely specifically on the Continuous Spontaneous Localization model \cite{Pearle1994}. As mentioned, the use of semi-classical gravity is regarded solely as an effective theory \footnote{As is well known, there are strong arguments indicating, although not definitely establishing, that semiclassical gravity might be unviable as a fundamental theory. See, for instance, \cite{Page, unviable-semicalsssical2}, while contrary opinions have been offered in \cite{unviable-semicalsssical7}.} rather than a fundamental framework for which the authors envision the need for a fully developed theory of quantum gravity. In this context, and in analogy with hydrodynamics, which is known to break down under certain conditions, we assume that the semi-classical gravity equations remain valid before and after a spontaneous collapse event, but not during the collapse itself, when the underlying assumptions of the approximation cease to be applicable. Significant advances have contributed to establishing a more rigorous mathematical basis for these ideas \cite{Juarez-Aubry:2019jon}. 

{It is worthwhile to note that as a byproduct of this approach, one obtains a natural account for a strong suppression of the generation of primordial gravity waves \cite{nobmodesbig} {with respect to the standard predictions of traditional inflationary models \cite{mukhanov2005}.}}

{For relatively large class of specific inflationary models the predictions concerning the spectrum of the tensor modes or primordial gravity waves are so close to the predictions concerning the scalar modes that consideration of the observed value for the amplitude and tilt of the latter together with the absence of detection of the former results in the conclusion that those specific models are ruled out \cite{Planck:2015jmz,planck2018like};{particularly the simplest ones. Although several concrete inflationary models remain viable despite these considerations; in particular, plateau-like inflaton potentials, which are considered unnatural by some theorists \cite{ijjas,ijjas2}. On the other hand, our approach, based on semiclassical gravity together with spontaneous reduction, generically leads to a strong suppression of the amplitude of tensor perturbations as compared to that of scalar perturbations. Thus, most natural inflationary potentials (e.g. $V \sim \phi^{2}$) remain, in principle, viable when analyzed within such a framework.}}

Once an alternative theory is proposed to address a problem, it should be carefully examined in connection with potential problematic implications. This was the main motivation for studying the eternal inflation problem in light of spontaneous collapse theories \cite{RL2023}. Cosmologists recognized the eternal inflation problem almost at the same time that inflation was incorporated into the standard cosmological model. The basic idea of this problem actually arises from a general misconception in physics regarding the notion of ``fluctuations''. We note that in principle there are at least three different notions that share the name ``fluctuations'': \textbf{I)} \textbf{The Quantum uncertainties}, which refer only to the indefiniteness of certain quantities, represented for instance, by the width of the Gaussian function in the ground state of the quantum harmonic oscillator; \textbf{II)} \textbf{The stochastic fluctuations}, associated with an ensemble in which each element has a definite value of said quantity; and say \textbf{III)} \textbf{The space-time variations}, which represent the changes within an extended system, of a particular quantity, from one point to another.

When deploying the standard considerations used in the treatment of inflation, an argument is made indicating that the ``zero mode'' of the inflaton field is subject to ``quantum fluctuations'' which, among other effects, creates a competition with the ``classical displacement'' described by the dynamical equations during the slow-roll period. This leads to two possibilities: \textbf{1)} Classical displacement dominates over the quantum fluctuations or \textbf{2)} Quantum fluctuations dominate over the classical displacement.
 
In the first case, we would not face a problem; however, the second leads to two types of situations 
\textbf{2a)} the zero mode of the inflaton moves backward (in the direction of an increasing potential), or \textbf{2b)} the mode moves forward (along a direction of decreasing potential). The issue is that when the inflaton is displaced in the direction of an increasing potential, the expansion rate increases, causing these regions to grow faster than regions where the inflaton moves in the opposite direction. Thus, over time, the regions with a higher expansion rate will cover a larger portion of the universe, and eventually dominate it as a whole, and therefore inflation would never end. This is the eternal inflation problem in a nutshell, see for instance \cite{lindeetef, guth2007}. As noted in the standard treatments, this problem is, in fact, tied to the erroneous identification of quantum uncertainties with stochastic fluctuations.

{{That issue, and its connection with the quantum measurement problem, has been emphasized by multiple authors (see, e.g., \cite{Norsen,Weinberg}), and several proposals aimed at sharpening the standard account can be found in the literature. In particular, there is an extensive program addressing the emergence of an effectively classical description during inflation via decoherence and related mechanisms \cite{Boddy:2014eba,Boddy:2016zkn}, as well as analyses based on backreaction and stochastic effects \cite{Brandenberger:2015rra,Markkanen:2016vrp}, and discussions stressing the role of vacuum squeezing in cosmology (see, e.g., \cite{Ashtekar:2020mdv}).}}

{{Nevertheless, as discussed in \cite{Berjon:2020vdv,Bengochea:2023dep} (see also \cite{Okon:2015fgr,Castagnino:2014cpa}), we do not consider these mechanisms to provide, by themselves, a fully satisfactory account of the specific issue emphasized here\footnote{{The issue is not in the calculations of the reduced density matrices that underlay decoherence or the analysis of the ``squeezzing'' of the state of quantum fields tied to the rapid cosmological expansion, but to the fact that these aspects by themselves can not justify the interpretation of the situation as involving any breakdown of the relevant symmetry. In fact as long as we take quantum theory to be more fundamental than the classical one, and thus accepting the classical description superveens on the quantum one, its simply a logical mistake to try to account for a breakdown of symmetry according to the former while admitting it is preserved according to the latter, see for a detailed discussion\cite{Bengochea:2023dep}.}}: the actual breaking of homogeneity and isotropy starting from an exactly homogeneous and isotropic quantum state evolving under symmetric dynamics. While decoherence can render interference terms in reduced density matrices negligible in an appropriate basis, and squeezing might motivate an effective classical description, neither step alone selects a single inhomogeneous realization (i.e. a definite outcome) or justifies interpreting the situation as an actual symmetry-breaking event.  In our view, this is precisely where an objective-collapse dynamics becomes relevant. Furthermore, the stochastic element often (and, in our view, incorrectly) attributed to quantum fluctuations is not provided by decoherence or by vacuum squeezing alone. }}

{{The latter point is particularly relevant in discussions of eternal inflation, where stochasticity plays a central role, but upon recognition that the traditional approaches provide no genuine source of stochastic behavior (unless some sort of ``measurement process'' is invoked), the problem would  seem to just disappear}}\footnote{It is worth noting that this misconception regarding fluctuations appears in many branches of physics, including high-energy physics. In a recent work \cite{Guerrero:2023wxw}, it was shown that the seeming discrepancy between the most recent bounds on the magnitude of the neutron electric dipole moment and the direct estimates of the corresponding quantum uncertainties can be traced to an erroneous interpretation of such quantum uncertainties as statistical fluctuations.}. However, as we also mentioned, the standard accounts lack a mechanism to break the homogeneity and isotropy of the quantum state, and thus one cannot possibly account for the emergence of cosmic structure. On the other hand, once we incorporate a spontaneous collapse theory into the picture, we face a context that does include stochastic fluctuations (affecting the dynamics of the inflaton field), a fact which now raises the possibility of the eternal inflation problem. 


 The work \cite{RL2023} explores a possible solution to this version of the eternal inflation problem, while retaining the spontaneous collapse account for the emergence of the seeds of cosmic structure. The approach incorporates a modification of the collapse rate obtained in \cite{Canate:2012nwv} with the functional form $\lambda=\lambda_{0}\left(\frac{k^{\alpha + 1}}{(\beta+k)^{\alpha}}\right)$. The aim is to provide a coherent account of the emergence of the seeds of cosmic structure, with an appropriate spectrum of scalar perturbations while avoiding the ``resurgence'' of the eternal inflation problem \footnote{This work is not the first to address this problem; in \cite{Leon:2017sru}, a study was carried out considering the collapse rate associated with the collapse of the field’s momentum.}. The preliminary analysis indicates a region of parameter space that seems, at first sight, as a valid one in both regards.

 In this manuscript, we aim to constrain the specific parameters of the model put forward in \cite{RL2023}, through a detailed numerical analysis comparing of the corresponding predictions directly with the observational data extracted from the most recent CMB studies. 

 In order to establish the model's viability, we will compare the fits corresponding to this model with those corresponding to the usual $\mathbf{{\Lambda}{{\textbf{CDM}}}}$ model, which we will refer to as the fiducial model. In order to achieve this, we must incorporate into the theory (as is done in \cite{piarunning}), the slow-roll parameters $\epsilon_{1}$ and $\epsilon_{2}$, and characterize those specifically, in terms of the Hubble flow functions \cite{terreroSR}. This reflects the need to identify clearly the cosmological parameters associated with the primordial spectrum, namely, the scalar amplitude $A_{s}$ and the scalar spectral index $n_{s}$; as both are dependent on $\epsilon_{1}$ and $\epsilon_{2}$.

The data analysis was carried out using a Markov chain Monte Carlo (MCMC) approach with the latest Planck 2018 \cite{Planck2018Results} likelihood code, adapting the CosmoMC software as necessary. The results indicate that they do not favor a specific value of $\alpha$, but they do constrain $\beta$. In order to address the eternal inflation problem, it was also necessary to review the inequality established in \cite{RL2023} and verify whether the choice of free parameters was consistent with the ranges imposed in that reference. Finally, we present the results from the exploration of the remaining cosmological parameters obtained when using the CSL model, in the regime where we have an avoidance of the eternal inflation condition.

The manuscript is structured as follows: Section 
\ref{Sec2} begins with a description contrasting the treatments of the primordial perturbations from the standard and the spontaneous collapse approaches. Then, the detailed quantum treatment is presented in the context of the CSL theory together with semiclassical gravity. To connect with the root of the problem, Section \ref{Sec:EternalCSL} is dedicated to the eternal inflation problem as viewed from within the spontaneous collapse framework. Afterward, in Section \ref{Sec4}, we present the analysis to assess the observational viability of our model. Finally, Section \ref{Sec5} is devoted to presenting the results and conclusions.

\section{Primordial perturbations: comparing the standard and the CSL model}\label{Sec2}

The main objective of this section is to present a comparison of the study of primordial perturbations of the power spectrum from two perspectives: using the standard approach and the Continuous Spontaneous Localization CSL theories. In the first subsection, we will present the standard idea of primordial perturbations, evidencing the conceptual problems at the end. {Then, in section 2.2, we presented the study of the primordial power spectrum in light of the CSL theory as has been study in \cite{Canate:2012nwv, piarunning}. In particular, we demonstrate how the interplay between semiclassical gravity and the CSL (in the context of semiclassical treatment of the inflationary universe) gives rise to the primordial inhomogeneities. To keep the natural flow of the lecture, we will add the main details of the CSL theory in the appendix \ref{apendiceRA}}.


\subsection{Standard treatment}\label{classical}

Following the standard inflationary approach, we take a single scalar field minimally coupled to gravity, with an appropriate potential. The action is:
\begin{equation}\label{action0}
\begin{split}
	S[\phi,g_{ab}] =& \int d^4x \sqrt{-g} \bigg[ \frac{1}{16 \pi G} R[g]\\
   & - \frac{1}{2}\nabla_a \phi \nabla_b \phi g^{ab} - V[\phi] \bigg]. 
\end{split}
\end{equation}
The background geometry is a flat FLRW spacetime described by the scale factor $a(t)$. In the matter sector, we model the scalar field as a homogeneous component plus small perturbations, $\phi(\x,t) = \phi_0(t) + \dphi(\x,t)$.

To characterize slow roll (SR) inflation, one introduces the Hubble flow functions $\epsilon_i$ (HFF) \cite{jmartinSR}. These are defined as is customary, 
\begin{equation}\label{defepsilonn}
	\epsilon_{n+1} \equiv \frac{d \ln \epsilon_n}{d N}, \qquad \epsilon_0 \equiv \frac{H_{\text{ini}}}{H},
\end{equation}
where $N \equiv \ln (a/a_{\text{ini}})$ is the number of e-folds from the start of inflation; $H \equiv \dot a/a$ is the Hubble parameter, and the dot denotes differentiation with respect to cosmic time $t$. Inflation occurs if $\epsilon_1 < 1$, and the slow roll approximation assumes these parameters remain small, $|\epsilon_n| \ll 1$. Moreover, since $d N = H dt$, one can derive another expression for the HFF involving the cosmic time:
\begin{equation}\label{dotepsilon}
	\dot \epsilon_n = H \epsilon_n \epsilon_{n+1}.
\end{equation}
Einstein field equations (EFE) obtained from action \eqref{action0}, yield the following dynamical equations of the (homogeneous) background:

\begin{equation}\label{friedamnnSR}
	H^2 = \frac{V}{ M_P^2 (3-\epsilon_1)},
\end{equation}
\begin{equation}\label{KGSR}
	3H \dot \phi_0 \left(1-\frac{\epsilon_1}{3} + \frac{\epsilon_2}{6} \right) = - \partial_{\phi_0} V,
\end{equation}
where $M_P^2 \equiv 1/(8 \pi G)$ is the reduced Planck mass. These equations hold exactly; however, imposing the slow-roll approximation $|\epsilon_i| \ll1$ with $i =1,2$, one obtains the slow-roll equations, 
\begin{equation}\label{friedamnnSR2}
	H^2 \simeq \frac{V}{3 M_P^2 },
\end{equation}
\begin{equation}\label{KGSR2}
	3H \dot \phi_0 \simeq - \partial_{\phi_0} V,
\end{equation}
 
At this point, we select a quadratic potential for the inflaton $V = m^2 \phi_0^2 /2$. This choice is motivated by two main considerations: First, the original model in Ref. \cite{RL2023} employed precisely this type of potential to tackle the issue of eternal inflation through a CSL-inspired framework. Second, unlike the standard slow-roll paradigm, in our model, polynomial potentials such as this one are not ruled out by current observations. Notably, in the context of CSL inflationary models under semiclassical gravity, a generic prediction is the significant suppression of the amplitude corresponding to primordial gravitational waves, irrespective of the specific potential assumed \cite{nobmodesshort,nobmodesbig}. 

Next, we move on to the perturbative sector of the theory. We begin by switching to conformal time coordinates, so the background metric can be written as $g_{\mu \nu}^{(0)} = a(\eta) \eta_{\mu \nu}$, where $\eta$ is the conformal cosmological time and $\eta_{\mu \nu}$ are the components of the Minkowski metric.

We choose the longitudinal gauge and concentrate on the first-order scalar perturbations. In this gauge, the line element of the metric is given by:
\begin{equation}
\begin{split}
	ds^2 = &a^2(\eta) \left[ - (1+2\Phi) d\eta^2 \right. \\
    & \left. (1-2 \Psi) \delta_{ij} dx^i dx^j \right] ,
\end{split}
\end{equation}

At first order in the perturbations, the components of Einstein's field equations (EFE), $\delta G_0^0 = 8 \pi G \delta T_0^0$, $\delta G_i^0 = 8 \pi G \delta T_i^0$, and $\delta G^i_j = 8 \pi G \delta T^i_j$, are written as
\begin{subequations}\label{EFEorden1}
\begin{equation}\label{00inf1} 
\begin{split}
	&\nabla^2 \Psi -3\mH(\mH\Phi + \Psi') = 4 \pi G \\
   & \times [-\phi_0'^2 \Phi + \phi_0' \dphi'
    + m^2 \phi_0 a^2 \dphi], 
\end{split}
\end{equation}

\begin{equation}\label{0iinf1} 
	\partial_i (\mH \Phi + \Psi') = 4 \pi G \partial_i ( \phi_0' \dphi), 
\end{equation}

\begin{equation}\label{ijinf1} 
    \begin{split}
        	&  [\Psi'' + \mH(2\Psi+\Phi)' + (2\mH' + \mH^2)\Phi \\
            &+ \frac{1}{2} \nabla^2 (\Phi - \Psi)] \delta^i_j \nonumber-\frac{1}{2} \partial^i \partial_j (\Phi - \Psi)  \\
	 &= 4 \pi G [\phi_0' \dphi' -\phi_0'^2 \Phi - m^2 \phi_0 a^2 \dphi]\delta^i_j. 
    \end{split}
\end{equation}
\end{subequations}

From Eq. \eqref{ijinf1} with $i \neq j$, we deduce that $\Psi = \Phi$. Hereafter, we adopt this result and refer to $\Psi$ as the Newtonian potential. Furthermore, in the longitudinal gauge, $\Psi$ corresponds to the curvature perturbation, i.e., the intrinsic spatial curvature on constant conformal time hypersurfaces.

Subtracting Eq. \eqref{00inf1} from Eq. \eqref{ijinf1}, and using Eq. \eqref{0iinf1} along with the equation of motion for the homogeneous scalar field (the exact one), $a^2 \partial_\phi V = -\phi_0'' - 2 \mH \phi_0'$, one obtains the evolution equation for the Newtonian potential: 

\begin{equation}\label{psievo} 
\begin{split}
	\Psi'' - \nabla^2 \Psi + 2 \left(\mH - \frac{\phi_0''}{\phi_0'} \right) \Psi' \\
    + 2 \left( \mH' - \frac{\mH \phi_0''}{\phi_0'} \right) \Psi = 0. 
\end{split}
\end{equation}

Using the change of variables
 \begin{equation}
	 \Psi = \left( \frac{\mH}{a^2 \theta} \right) u; \qquad \theta \equiv \frac{ \mH }{a \phi_0'},
	 \end{equation} the above evolution equation becomes \begin{equation}
	  u'' - \nabla^2 u - \frac{\theta''}{\theta} u = 0. \end{equation}

In Fourier space,\footnote{We define the Fourier transform of a function $f(\x,\eta)$ as 
	$$f(\x,\eta) = \frac{1}{(2 \pi)^{3/2}} \int_{ \mathbb{R}^3} d^3k \: e^{i \nk \cdot \x} f_\nk (\eta) $$ } the solution to the above differential equation in the super-Hubble limit $k \ll \mH$ is given by
    
	 \begin{equation} 
     \begin{split}
     u_k (\eta) \simeq& C_G (k) \theta(\eta) \int^\eta \frac{d\tilde \eta}{\theta^2(\tilde \eta)}\\
     &+ C_D(k) \theta(\eta),
     \end{split}
	 \end{equation}
     
     where the coefficients $C_G(k)$ and $C_D(k)$ are constants (functions of $k$, independent of $\eta$) determined by the initial conditions. Thus, the scalar perturbation $\Psi_k$ is

\begin{equation}\label{solPsi} 
\begin{split}
\Psi_k (\eta) \simeq& C_G (k) \frac{\mH}{a^2} \int^\eta \frac{d\tilde \eta}{\theta^2(\tilde \eta)} \\
&+ C_D(k) \frac{\mH}{a^2}. 
\end{split}
\end{equation}

During inflation, the second term in Eq. \eqref{solPsi}, known as the decaying mode, can be neglected. The first term can be expanded using the HFF by employing the exact relation\footnote{Here we have used the exact relation
	\begin{equation*}
		\phi_0'^2 = 2 \epsilon_1 \mathcal{H}^2 M_P^2,
	\end{equation*}
	where $M_P^2 = (8 \pi G)^{-1}$ is the reduced Planck mass); this expression is obtained from Friedmann's equations, with matter modeled as a single scalar field. } \begin{equation}\label{eqeqeq} 
\theta^{-2}	 = 2 a^2 \epsilon_1 M_P^2, 
	\end{equation} 
	where $M_P^2 = (8 \pi G)^{-1}$ is the reduced Planck mass. Substituting this into Eq. \eqref{solPsi} and integrating by parts (making use of definitions \eqref{dotepsilon}), the solution expanded up to second order in the HFF is given by 
	\begin{equation}\label{solPsi2}
    \begin{split}
		 \Psi_k (\eta) &= C_G (k) [ \epsilon_1 + (\epsilon_1^2 + \epsilon_1 \epsilon_2)] \\
         &+ \mathcal{O}(\epsilon_n^3). 
       \end{split}  
         \end{equation}

Equation \eqref{solPsi2} quantitatively establishes the well-known result \cite{mukhanov92,mukhanov2005} that $\Psi_k$ remains approximately constant (independent of $\eta$) for super-Hubble modes.

In order to compare the theoretical predictions with observational data, one relies on the use of the so-called \textit{comoving curvature perturbation} $\mR$.\footnote{For example, the \textit{Planck} collaboration uses the scalar power spectrum of $\mR$ in order to constrain inflationary parameters, such as the amplitude $A_s$ and the scalar spectral index $n_s$ \cite{Planck2018_inflation}.} This variable can be expressed in terms of $\Psi$ and $\dphi$ as 
\begin{equation}\label{Rexacta}
	\mR \equiv \Psi + \frac{\mH}{\phi_0'} \dphi.
\end{equation}

Switching to Fourier space, and for the observational modes of interest, taking the solution \eqref{solPsi2} at the lowest order in the HFF yields
\begin{equation}\label{masterR0}
\begin{split}
	\mR_\nk (\eta) &\simeq C_G (k) \epsilon_1 + \frac{1}{\sqrt{2 \epsilon_1} M_P} \dphi_\nk (\eta)\\
    &\simeq \frac{1}{\sqrt{2 \epsilon_1} M_P} \dphi_\nk (\eta),
\end{split}
\end{equation}
where in the last equality, we have used that during inflation $\epsilon_1 \ll 1$. 

From this point, the standard treatment involves quantizing $\mathcal{R}$, which, as indicated in (\ref{Rexacta}), requires a simultaneous quantization of both the metric and the field perturbations. Afterward, one selects the so-called Bunch-Davies vacuum state and compute the corresponding quantum two-point correlation function:

\begin{equation}\label{quspect}
    \bra{0}|\mathcal{R}_{k} (\eta)\mathcal{R}_{k'} (\eta)|0\ket,
\end{equation}  
This step, usually taken with no or almost no discussion (see \cite{Weinberg:2008zzc} for an exception), is both innocent-looking and at the same time highly misleading. In the setting of spatially dependent quantities (i.e. when considering $\bra{0}|\mathcal{R}(x, \eta)\mathcal{R}(x', \eta)|0\ket $), one is interpreting a quantum object meant to characterize quantum entanglement with a statistical description meant to describe a space-temporal variation of a certain quantity. By passing to the Fourier description one ends up identifying, what is essentially a measure of a quantum uncertainty, $( \Delta \hat O ) ^2 = \bra{0}|\hat O^2 |0\ket - \bra{0}|\hat O|0\ket^2 = \bra{0}|\hat O^2 |0\ket $, with that of a stochastic fluctuation of classical valued quantity. The reader could just consider if, when examining, say, an EPR-B type experiment, such identifications can be considered as justifiable in the context in which there is no measurement (or if there are measurements which take place afterward). For a more exhaustive discussion, the reader is referred to works such as \cite{Sudarsky:2009za, Berjon:2020vdv}.

 From this two-point function, the power spectrum is read using the relation 
\begin{equation}
 \bra{0}|\mathcal{R}_{k} (\eta)\mathcal{R}_{k'} (\eta)|0\ket
 = \int d^{3}k \, e^{i\vec{k}\cdot(\vec{x}-\vec{y})} \, \bra\mathcal{R}(x)\mathcal{R}(y)\ket.
\end{equation} 
The resulting spectrum is $\mathcal{P}(k) \sim k^{-3}$, up to small corrections characterized by the slow-roll parameters $\epsilon_1, \epsilon_2$. 

Here we can see, on the one hand, the standard theory from which it is often claimed that inflation reproduces the primordial structure, and at the same time, one of the most common conceptual issues in early-universe cosmology: identifying the two-point function (or related quantum uncertainties) with the average over possible classical universes. In fact, another drawback of such an approach is that in order to describe individually one of such possible universes, one must introduce some {\it ``ad hoc'' } manipulations, often including stochastic features, whose origin remains completely unclear (see, for instance, the discussion on page 246 of {\cite{Peter:2013avv}}).

\subsection{Semiclassical treatment: The CSL mechanism and the primordial power spectrum}\label{sec_22}


{In this section, we summarize the main ideas of modifying the standard treatment based on CSL theory and semi-classical gravity \cite{Diez:2012, Canate:2012nwv, piarunning2}. The details of CSL theory are covered in the Appendix \ref{apendiceRA}. CSL theories introduce a change in the dynamics introducing a stochastic term in the evolution equation of the state, this leads to the replacement of the classical field $\dphi_\nk$ with the quantum expectation value (also classical) $\bra \hat \dphi_\nk \ket$} \footnote{In principle, when relying on the semi-classical gravity framework, one should simply and directly replace the classical energy momentum tensor with the corresponding expectation value. However, as shown in \cite{Diez:2012}, the fact that the creation and annihilation operators associated with the zero mode, as well as those corresponding to the spatially dependent modes, commute, together with the assumption that the modes are unentangled (to lowest order), indicates that we may proceed to a treatment in which the fields themselves are replaced by their expectation values. Note, by the way, that for the Bunch Davies vacuum we have $\bra \hat \dphi_\nk \ket=0$ and it is only due to the collapse dynamics that the state of the quantum field is modified and this quantity acquires a non-vanishing value.}. As a result of all this, we have:
 
\begin{equation}\label{masterR}
	\mR_\nk (\eta) \simeq \frac{1}{\sqrt{2 \epsilon_1} M_P} \bra \hat \dphi_\nk (\eta) \ket,
\end{equation}

which is an equation with no analog in the standard treatments (the corresponding result would indicate $\mR_\nk (\eta) =0 $, which would be clearly rather problematic). 

The scalar power spectrum of $\mR_\nk$ is now defined by
\begin{equation}\label{PSdef}
	\overline{\mR_{\nk}\mR^*_{\nq}} \equiv \frac{2 \pi^2}{k^3} P (k) \delta(\nk-\nq),
\end{equation} 
where $P (k)$ is the \textit{dimensionless} power spectrum. The bar appearing in Eq. \eqref{PSdef} denotes an ensemble average over possible realizations of the stochastic field $\mR_{\nk}$ corresponding now clearly to an average over possible universes resulting from the stochastic features of the dynamics of the spontaneous collapse theory. Thus in our approach, each realization corresponds to a particular outcome of the stochastic process characterizing the spontaneous collapse mechanism\footnote{As noted in the standard approach, it is common to find the misconception of treating at the same level the quantum correlations between two operators and the ensemble over distinct possible realizations, which of course only refer to the situation after a corresponding measurement has taken place, something that would be unjustifiable in traditional quantum mechanical treatments}. Therefore, from Eqs. \eqref{masterR} and \eqref{PSdef}, the scalar power spectrum can be identified as
\begin{equation}\label{masterPS}
	P (k) \delta(\nk-\nq) = \frac{k^3}{4 \pi^2 M_P^2 \epsilon_1 } \overline{\bra \hat \dphi_\nk \ket \bra \hat \dphi_\nq \ket^*}.
\end{equation}

The quantity $\overline{\bra \hat \dphi_\nk \ket \bra \hat \dphi_\nq \ket^*}$ must be evaluated in the super-Hubble regime $k \ll a H$. In the next section, we will focus on this evaluation.

We now focus on the quantum description of the perturbations. Our analysis is built upon the quantum field theory (QFT) of the perturbation field $\dphi(\x,\eta)$, living in a background spacetime of the quasi–de Sitter type. By expanding the action \eqref{action0} up to second order in the perturbations, one can obtain the action corresponding to the matter perturbations. Since our approach employs the semiclassical gravity framework, only the matter degrees of freedom are quantized; however we keep terms at linear order in the HFF. Moreover, introducing the rescaled variable $y = a \dphi$, the second-order action becomes $\delta^{(2)} S = \int d\eta \int_{\mathbb{R}^{3+}} d^3\nk \mathcal{L}$, 

\begin{equation}\label{action2}
\begin{split}
	\mathcal{L} &= y_{\nk}' y_{\nk}^{\star '} - k^2 y_{\nk} y_{\nk}^{\star}\\
	&-\frac{z'}{z} \left( y_{\nk} y_{\nk}^{\star '} + y_{\nk}' y_{\nk}^{\star}
	\right) + \left( \frac{z'}{z} \right)^2 y_{\nk} y_{\nk}^{\star},
\end{split}
\end{equation}
with $y_{\nk}(\eta)$ the Fourier modes associated to the field $y(\x,\eta)$ and $z \equiv a \sqrt{\epsilon_1} $. 

The CSL model is based on a non-unitary modification to the Schr\"odinger equation; consequently, it is suitable to perform the quantization of the fields in the Schr\"odinger picture, where the relevant physical objects are the Hamiltonian and the wave functional. 

We first define the canonical conjugated momentum associated to $y_{\nk}$, this is $p_{\nk} \equiv
\partial \mathcal{L}/ \partial y_{\nk}^{\star '}$, or explicitly $p_{\nk} = y_{\nk}'- y_\nk z'/z$. The Hamiltonian associated to Lagrangian $\mathcal{L}$, can be found as $\mH = \int_{\mathbb{R}^{3+}} d^3k \: (y^{*'}_\nk p_\nk + y^{'}_\nk p^*_\nk  ) - \mathcal{L}$. Therefore, $\mH =\frac{1}{2}\int_{\mathbb{R}^{3+}} d^3 \nk\: (H_{\nk}^\text{R} +H_{\nk}^\text{I})$, where
\begin{equation}\label{hamRI}
\begin{split}
H^{R,I}_\mathbf{k } = &\left( p_\mathbf{k }^\RI \right)^2 + k^2 \left( y_\mathbf{k }^\RI \right)^2 \\
&+ \frac{z'}{z} \left( y_\mathbf{k }^\RI p_\mathbf{k }^\RI + 
p_\mathbf{k }^\RI y_\mathbf{k }^\RI \right) ,
\end{split}
\end{equation}
The superscripts R,I denote the real and imaginary parts of $y_\nk$ and $p_\nk$, i.e. 
\begin{equation}
	y_\nk \equiv \frac{1}{\sqrt{2}} (y_\nk^\text{R} + i y_\nk^\text{I} ), \qquad p_\nk \equiv \frac{1}{\sqrt{2}} (p_\nk^\text{R} + i p_\nk^\text{I} ),
\end{equation}

Our next step is to describe the quantization of the system. We will assume that the collapse is somehow analogous to an imprecise measurement of the field and momentum operators. Therefore, their real and imaginary parts are completely Hermitian, and thus, qualify as reasonable observables (and thus are susceptible to their treatment as collapse inducing operators, according to the general structure of SCL-theories). We promote then ${y}_{\nk}, {p}_{\nk}$ to quantum operators satisfying the canonical commutation relations $ 	[\hat{y}_{\nk}, \hat{p}_{\nq}] = i\delta (\nk-\nq) $.

In the Schrödinger picture, the state of the system is characterized by the wave functional $\Phi[y(\x,\eta)]$. In Fourier space, the theory is still free in the sense that it does not contain terms with power higher than two in the Lagrangian; therefore, the wave functional can be factorized into contributions from individual modes as $\Phi[y(\x,\eta)] = \Pi_{\nk}\Phi_{\nk}^R(y_{\nk}^R)\Phi_{\nk}^I(y_{\nk}^I)$. From this point forward, each mode will be treated independently. In the field representation, the operators would take the form:
\begin{equation}
	\hat y_\nk^\RI \Phi = y_\nk^\RI \Phi, \qquad \hat p_\nk^\RI \Phi = -i \frac{\partial \Phi }{\partial y_\nk^\RI}.
\end{equation}

As a result of each mode evolving independently, the Schrödinger equation can be decomposed into an infinite set of ordinary differential equations, one for each individual mode $\Phi_\nk$. The evolution consists of two parts: (i) the traditional one given by the ``free'' Hamiltonian plus (ii) the CSL evolution term. Specifically, the free part is 
\begin{equation}\label{schrodeq}
	i \frac{\partial \Phi_\nk^\RI }{ \partial \eta} = \hat H^{(0)\RI}_\nk \Phi_\nk^\RI,
\end{equation}
with the free Hamiltonian explicitly given in Eq. \eqref{hamRI}.

Before discussing the evolution provided by the CSL mechanism, let us address the initial conditions next. The standard assumption is that at very early conformal times, $\tau \to -\infty$, each mode is found in its ground state, which is a Gaussian, centered at zero, and characterized by a specific spread. This initial state is commonly known as the Bunch-Davies (BD) vacuum (with the conformal time $\eta$ in the interval $[\tau,\eta_0)$  with $ \eta_0 < 0$ ) .

Given that the initial quantum state is Gaussian, and considering the Hamiltonian is quadratic\footnote{The modification introduced by the CSL mechanism is also quadratic in the field variables.} in the operators $\hat y_\nk^\text{R,I}$ and $ \hat p_\nk^\text{R,I}$, the wave functional maintains a Gaussian form in the field representation at all times:
\begin{equation}\label{psionday}
\begin{split}
	\Phi^{R,I}(\eta,y_{\nk}^\RI)& = \exp[- A_{k}(\eta)(y_{\nk}^\RI)^2  \\+
	B_{k}(\eta)y_{\nk}^\RI
    &+ C_{k}(\eta)].
\end{split}
\end{equation}

Thus, the evolution of the wave functional is subject to the following initial conditions:
\begin{equation}
	A_k (\tau ) = \frac{k}{2}, \hspace{4pt} B_k (\tau ) = C_k(\tau )= 0.
\end{equation}
These initial conditions define the BD vacuum, a perfectly homogeneous and isotropic state consistent with the notion of vacuum states in QFT.

To incorporate the CSL model during inflation, we follow the methodology originally proposed in \cite{pedrocsl} within the inflationary context, and which is also the same model adopted in Ref. \cite{RL2023} to address eternal inflation. In summary, it was demonstrated that by choosing suitable field collapse operators and employing the corresponding CSL evolution equation, one can achieve a ``collapse'' for the relevant operators associated with the Fourier components of the field. Additionally, it was noted that the collapse mechanism affects independently each mode of the field, without introducing mixing between different modes.

In this framework, the CSL evolution of the state vector corresponding to each mode of the quantum field is given by:

\begin{equation} \label{CSLevolution}
\begin{split}
       & |\Phi_{\nk}^{\textrm{R,I}}, \eta \ket = \hat{T}\exp \bigg\{-\int_{\tau}^{\eta}d\eta' \bigg[ i \hat{H}_{\nk}^{(0) \RI}\\
       &+\frac{1}{4 \lambda_k}(\mathcal{W}_{\nk}^{\RI}(\eta) - 2 \lambda_k
	\hat{y}_{\nk}^{\textrm{R,I}})^2 \bigg]  \bigg\},
\end{split}
\end{equation}
where $\hat T$ represents the time-ordering operator, and $\tau$ is the conformal time marking the onset of inflation. The stochastic field $\mathcal{W}_\nk= \mathcal{W}_{\nk}^{\text{R}} + i \mathcal{W}_{\nk}^{\text{I}}$ explicitly depends on the mode $\nk$ and conformal time. Since we apply CSL dynamics independently to each field mode, it is natural to introduce a distinct stochastic function for each mode. Therefore, the stochastic field $\mathcal{W}_\nk (\eta)$ can be viewed as the Fourier transform of a stochastic spacetime field $\mathcal{W}(\x,\eta)$. The probability distribution for the stochastic field is governed by the second fundamental CSL equation—the Probability Rule—which takes the form:
\begin{equation}\label{cslprobabF}
\begin{split}
	P(\mathcal{W}_{\nk}^{\RI}) d\mathcal{W}_{\nk}^{\RI} =  \bra \Phi_{\nk}^{\RI} , \eta | \Phi_{\nk}^{\RI}, \eta \ket  \\ \times \prod_{\eta'=\tau}^{\eta-d\eta} \frac{ d \mathcal{W}_{\nk}(\eta')^{\RI}}{\sqrt{2 \pi \lambda_k/d\eta}}.
\end{split}
\end{equation}

As seen in the CSL evolution equation \eqref{CSLevolution}, the operator $\hat{y}_\nk^\RI$ has been chosen as the generator of the collapse process. This choice implies that the CSL dynamics will drive the initial quantum state toward an eigenstate of $\hat{y}_\nk^\RI$. The motivation for selecting this particular collapse operator lies in the central role played by the field $\dphi$ in the eternal inflation scenario, as well as in its role as the source of the comoving curvature perturbation $\mR$, as expressed in Eq. \eqref{masterR}.

In terms of the Real and Imaginary parts of $\hat y_\nk$, the ensemble average appearing in the expression for the primordial power spectrum, Eq. \eqref{masterPS}, is
\begin{equation}\label{igualdad00}
	\overline{ \bra \hat y_{\nk} \ket \bra \hat y_{\nq} \ket^* } = \frac{1}{2} (\overline{\bra \hat y_{\nk}^\textrm{R} \ket^2} + \overline{\bra \hat y_{\nk}^\textrm{I} \ket^2}) \delta (\nk-\nq).
\end{equation}
Furthermore, $\overline{ \bra \hat y_{\nk}^\textrm{R} \ket^2 }= \overline{\bra \hat y_{\nk}^\textrm{I} \ket^2}$, thus we will omit the indexes R,I from now on.

Using the main equations of the CSL model, Eqs. \eqref{CSLevolution} and \eqref{cslprobabF}, we can obtain (see Refs. \cite{pedrocsl,RL2023} for details):
\begin{equation}\label{igualdad0}
	\overline{\bra \hat y_{\nk} \ket^2} = \overline{\bra \hat y_{\nk}^2 \ket} - \frac{1}{4 \textrm{Re} [A_k(\eta)]}.
\end{equation}
Substituting Eqs. \eqref{igualdad00} and \eqref{igualdad0} into Eq. \eqref{masterPS}, we find our prediction of the primordial power spectrum:
\begin{equation}\label{masterPS2}
\begin{split}
	P(k) =& \frac{k^3}{4 \pi^2 M_P^2 a^2 \epsilon_1^2}\\
   &\times \left( \overline{\bra \hat y_{\nk}^2 \ket} - \frac{1}{4 \textrm{Re} [A_k(\eta)]} \right).
\end{split}
\end{equation}
The terms $ \overline{\bra \hat y_{\nk}^2\ket }$ and $(\textrm{Re} [A_k(\eta)])^{-1} $, appearing in the later expression of the power spectrum, can be obtained by solving explicitly the corresponding CSL equations. We have included such a procedure in Appendix \ref{apendiceA} for the interested reader.

After evaluating each term on the right-hand side of Eq. \eqref{masterPS2}, the resulting expression for the power spectrum in our model takes the form
\begin{equation}\label{PSfinal}
	P(k) = A_s \left( \frac{k}{k_*} \right)^{n_s-1} C(k),
\end{equation}
where
\begin{equation}
	A_s \equiv \frac{H_*}{8 \pi^2 M_P^2 \epsilon_{1 *}}, \qquad n_s = 1 - 2 \epsilon_{1 *} - \epsilon_{2 *},
\end{equation}
with $A_s$ denoting the amplitude of the spectrum, and $n_s$ representing the spectral index. The expression for $n_s$ in terms of the HFF remains the same as in standard treatments (see Ref. \cite{piarunning} for the derivation up to second order in the HFF). The function $C(k)$, introduced by the CSL mechanism, is defined as

\begin{equation}\label{defCk}
\begin{split}
     C(k)& \equiv  \frac{\lambda_k |\tau|}{k} + \frac{3}{2} \frac{\lambda_k}{k^2} \sin (2k|\tau|)  \\
	&+1 - \bigg[ \frac{2 \lambda_k}{k^2} \left( \frac{k}{k_*} \right)^{n_s-2}\\
    &+ \zeta_k^{4-n_s} \cos\bigg((4 - n_s) \theta_k\bigg) \bigg]^{-1} ,   
\end{split}
\end{equation}

with
\begin{equation}\label{defzetatheta}
\begin{split}
	\zeta_k \equiv \left(1 + \frac{4 \lambda_k^2}{k^4}\right)^{1/4},\\
    \theta_k \equiv - \frac{1}{2} \arctan \left(\frac{2 \lambda_k}{k^2}\right).
\end{split}
\end{equation}
The subscript $_\star$ indicates that such a quantity is evaluated at the time of ``horizon crossing'' of the pivot mode $k_\star$, i.e. when $-k_\star \eta_\star = 1$

It is worth noting that the first two terms dominate the spectrum; when $\lambda_k = 0$, they vanish immediately, whereas the remaining two terms cancel each other. The latter is because, in this limit, we also have $\zeta_k = 1$ and $\theta_k = 0$, which leads to $C(k) = 0$ and consequently $P(k) = 0$. This result is consistent with our model, where the emergence of metric perturbations is attributed to the collapse of the wave function induced by the CSL mechanism. Thus, in the absence of collapse--i.e., if the quantum state remains in the homogeneous and isotropic vacuum--no primordial perturbations are generated.

Another point worth mentioning is that if we set $n_s = 1$ exactly in \eqref{defCk}, we have verified that the resulting expression matches the one obtained in Refs. \cite{pedrocsl,RL2023}, where a perfectly scale-invariant power spectrum was derived by approximating the background spacetime as nearly de Sitter, i.e., by assuming $a(\eta) \simeq -1/H \eta$ and $\epsilon_1 =$ const. In fact, this approximation was also employed when addressing the problem of eternal inflation in \cite{RL2023}. The point is that in addressing the eternal inflation issue, the de Sitter approximation remains sufficient and no significant modifications emerge from the inclusion of the next-to-leading order in the HFF (in contrast with the prediction for the primordial power spectrum as done in the present section). Consequently, the parametrization for $\lambda_k$ found in Ref. \cite{RL2023}, used in dealing with the eternal inflation problem, remains fully applicable, as we will see when we treat this subject in the next section. 

\section{Eternal inflation within the CSL framework}
\label{Sec:EternalCSL}
As we pointed out in the preceding section, in order to provide a satisfactory explanation for the primordial cosmic structure, it is necessary to introduce collapse as the mechanism that breaks the Bunch–Davies symmetries (homogeneity and isotropy) \cite{pedrocsl}. However, the inclusion of a stochastic term (associated with the collapse) in the evolution equation inherently adds stochastic fluctuations, which opens the possibility for eternal inflation, a point that will be
briefly addressed in what follows.

The eternal inflation scenario can be analyzed within the slow-roll regime. The inflaton zero mode $\phi_0$ gradually rolls toward the potential minimum. Over one Hubble time, in addition to the classical displacement $\Delta_{\rm class}\phi_0 \equiv \phi_0' \Delta \eta$, as dictated by the slow-roll dynamics, the field $\phi_0$ also undergoes stochastic fluctuations\footnote{In the standard theory, it is common to find confusion between quantum and stochastic fluctuations, often treated as synonyms. Nonetheless, the former refers merely to the indeterminacy of certain quantities,  whereas the latter describes the variation of a value across an ensemble or a large collection of outcomes. In our approach, however, these stochastic fluctuations originate from the collapse process.}, denoted $\Delta_{\rm sto}\phi_0$. Due to its stochastic character, two distinct possibilities arise:

\begin{itemize}
	\item[a)] The classical displacement of the field dominates over the stochastic fluctuations,
	\begin{equation}\label{noet}
		\Delta_{\text{sto}}\phi_{0}<\Delta_{\text{class}}\phi_{0},
	\end{equation}
	this condition can be understood as the non-eternal inflation condition, so the evolution of the universe can follow the standard cosmological model.
	\item[b)] Stochastic fluctuations are greater than the classical displacement:
	\begin{equation}
		\Delta_{\text{sto}}\phi_{0}>\Delta_{\text{class}}\phi_{0}.
	\end{equation}
\end{itemize}
The dominance of stochastic fluctuations over classical ones gives rise to two outcomes: (1b) Regions where the inflaton zero mode rolls downward toward the potential minimum, inducing an expansion driven by an effective cosmological constant, determined by a ``small'' potential; (2b) Regions where the zero mode rolls upward acquire a larger effective cosmological constant. As time progresses, the latter regions come to dominate the volume of the Universe, producing the well-known phenomenon of eternal inflation.

In Section \ref{sec_22}, we have mentioned that CSL theory requires the choice of a collapse operator. As shown in various works on the subject, \cite{pedrocsl,piarunning,piarunning2}, this choice is strongly connected with the functional form and value of the collapse parameter. In the present paper, as well as in Ref. \cite{RL2023} (where the eternal inflationary for this CSL model was investigated), we have selected the inflaton field itself as the collapse operator. For this choice, the simplest form of $\lambda_k$ is \cite{piarunning,piarunning2}:
\begin{equation}\label{lambdak0}
	\lambda_k = \lambda_0 k
\end{equation}
where $k$ is the wave number of each mode, and $\lambda_0$ can be related to the universal CSL rate parameter, which has units of [Time]$^{-1}$. Thus $\lambda_0^{-1}$ fixes the spontaneous localization time scale for the wave function associated to each mode of the field. The form \ref{lambdak0} will be analyzed in the next Section, but we anticipate that it should yield a primordial power spectrum which, when adjusted to the current CMB data, should be very close to the one obtained in the standard treatments.

On the other hand and looking at the situation from the perspective of the eternal inflation problem, it might seem that, as inflation is driven by the inflaton's zero mode ($k=0$), and for such mode $\lambda_k =0$, the issue simply disappears as a vanishing the collapse rate 
implies that no stochastic fluctuations would be affecting that sector. However, the possibility of eternal inflation reemerges when we note the need to consider modes associated with scales large enough that effectively mimic the zero mode. In particular, the problematic modes are those which contribute to the Newtonian potential $\Psi$ can be taken as depending just on conformal time, i.e. modes that 
are in practice essentially spatially constant. 
The point is that the contribution to $\Psi$ in this case leads to a line element which can be described by

\begin{equation}
\begin{split}
  	ds^{2} = a^2 (\eta)& [ - (1 + 2 \Psi(\eta)) {d \eta}^2 \\
    &+ (1 - 2 \Psi(\eta)) \delta_{ij} d x^i d x^j ] \\
	&= \tilde a^2 (\tilde \eta) [ - d {\tilde \eta}^2 + \delta_{ij} d x^i d x^j ],  
\end{split}
\end{equation}

where $d {\tilde \eta} = \bigl[(1 + 2 \Psi(\eta))(1 - 2\Psi(\eta))^{-1}\bigr]^{1/2} d \eta$ and $\tilde a = a (1 - 2 \Psi)^{1/2}$. In other words, it is essential to consider modes with $k \to 0$ whose Newtonian potential remains approximately constant in space, as they can produce effects analogous to those of the zero mode, thereby influencing the evolution of the causally connected region. 

In Ref. \cite{RL2023}, it was argued that a modification of Eq. \eqref{lambdak0},  capable of preventing the long-wavelength modes discussed above from becoming dominant in in $\Delta_{\rm sto}\phi_0$ and leading to eternal inflation, is given by
\begin{equation}\label{lambdaketernal}
	\lambda_k = \lambda_0 \frac{k^{\alpha+1}}{(\beta + k)^\alpha},
\end{equation}
where two new parameters were introduced: $\alpha$, which is dimensionless and positive, and $\beta$, which carries the same units as $k$. If either $\alpha=0$ or $\beta=0$, this expression reduces to the form given in Eq. \eqref{lambdak0}.

Next, we will explore in detail the phenomenological viability of that proposal.

\subsection{Observational viability of the CSL model without restrictions}
\label{Sec:CSLobs}




In the previous sections, we presented both our general framework and the particular parametrization of the collapse rate we want to explore.
The proposal introduced the two parameters $\alpha$ and $\beta$, whose values will be explored in search of consistency with observations. 
From now on, we refer to the {\em fiducial model} as the standard $\Lambda$CDM cosmology, which will serve as a baseline for analyzing the specific features of the CSL inflationary scenario.

At this point, it is useful to discuss some relevant numerical values for the main physical quantities appearing in the model. We assume that inflation lasts 65 e-folds approximately, so $a_{\rm end} = a_{\rm ini} e^{65}$. Additionally, we also assume that inflation ends at a temperature of $10^{-4} \text{M}_P$; comparing this scale with the CMB temperature today $T_{\rm cmb} \simeq 0.23$ meV, we have $a_0/a_{\rm end} = T_{\rm end}/T_{\rm cmb} = 10^{-27}$. Therefore, using the previous estimates, we obtain an approximate value for the scale factor at the beginning of inflation $a_{\rm ini}/a_0 = 10^{-56}$. Finally, using the form of the scale factor $a(\eta) = -1/(H \eta)$ with $H\simeq 10^{-9} \text{M}_P$ approximately constant during inflation and the normalization $a_0=1$, we can obtain an estimate for the conformal time at the beginning of inflation $-\tau = 1/H a_{\rm ini}$, which yields $-\tau \simeq 10^{8}$ Mpc. Regarding $\lambda_0$, during the computation of $P(k)$ in Appendix \ref{apendiceA}, we have used the approximation $\lambda_0 |\tau| \gg 1$. Hence, we fix $\lambda_0 = 10^{-5}$ Mpc$^{-1}$ (or in international units $\lambda_0 = 10^{-19}$ s$^{-1}$), in this way $\lambda_0 |\tau| = 10^3$.

\begin{figure}[h]
    \centering
    \includegraphics[angle=270,width=0.5\textwidth]{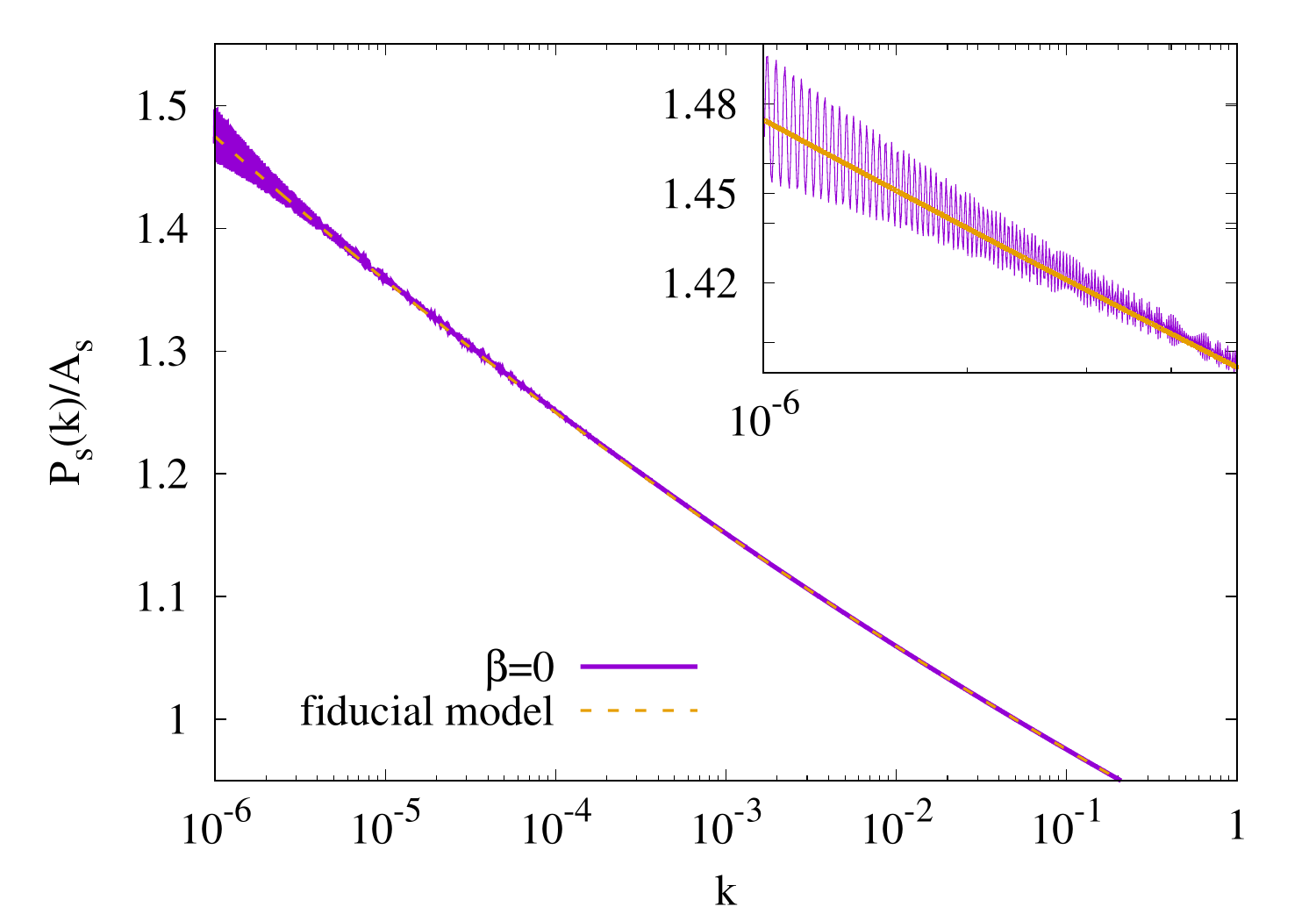}
    \caption{Primordial power spectrum with $\beta=0$ compared to standard $\Lambda$CDM (fiducial) model in dashed line. According to our proposal, selecting a null value for $\beta$ parameter recovers the functional form of the standard power law, except for the oscillations at low $k$.}
    \label{fig:Bcero}
  \end{figure}

Figure \ref{fig:Bcero} plots Eq. \eqref{PSfinal} normalized by $A_s$ in the case $\beta=0$ [so $\lambda_k = \lambda_0 k$ from Eq. \eqref{lambdaketernal}], together with the standard $P(k)$ of the fiducial model. The CSL power spectrum for $\beta=0$ reproduces the usual power-law shape at leading order; nevertheless, small-scale oscillatory features persist. These oscillations are a characteristic signature of the CSL collapse mechanism, but are of sufficiently low amplitude that they remain unobservable in the angular power spectrum. Hence, the differences between our proposal, when $\beta=0$, and the standard scenario arise purely at the theoretical and conceptual levels.

The oscillations observed in the power spectrum, Fig.~\ref{fig:Bcero}, are a characteristic feature of the CSL model and are absent in the standard framework. They arise from the way we relate the primordial spectrum to the (squared) quantum expectation value of each field mode, i.e.\ $P(k) \propto \langle \dphi_k \rangle^{2}$; note that this is not the quantum uncertainty in the vacuum, $\langle O^{2} \rangle$, which is the quantity usually associated with the power spectrum in the traditional treatments of inflation. From the computational point of view, the oscillations originate from solving the CSL equations with the choice of initial conditions for the usual modes in BD vacuum.\footnote{ As shown in \ref{apendiceA}, the initial condition of the CSL model enters through the constants $C_1$ and $C_2$ in Eq. \eqref{Q} leading to Eq. \eqref{Qaprox}. In particular, because the Bunch-Davies (BD) vacuum is imposed together with the CSL evolution, the constants $C_{1}$ and $C_{2}$ do not vanish, so the oscillatory terms in \eqref{Qaprox} persist. This behavior contrasts with standard treatments of inflation, where one would have  $C_{1}=C_{2}=0$ \cite{Canate:2012nwv}, and where the oscillatory contributions are absent.  } Furthermore, still for $\beta = 0$, the function $C(k)$ in \eqref{defCk}, which encodes the CSL contribution to the power spectrum, can be approximated as,
\begin{equation}\label{Ckapp}
    C(k) \simeq \lambda_0 |\tau| \left[1 + 3\frac{\sin (2k|\tau|)}{2k|\tau|}  \right].
\end{equation}
Therefore, the maximal amplitude of the oscillatory term within the observable $k$ range occurs around $k \in [10^{-6},10^{-5}]$ Mpc$^{-1}$. Taking into account that $|\tau| = 10^8$ Mpc, we can estimate the amplitude of the oscillations in \eqref{Ckapp} as $(k |\tau|)^{-1} \simeq [10^{-2}, 10^{-3}]$, consistent with what we see in the zoomed region in Fig.~\ref{fig:Bcero}.

Physically, the oscillations can be understood by analogy with the quantum harmonic oscillator. In its ground state, the wavefunction is a Gaussian centered at zero with finite width. The peak of the Gaussian represents the expectation value $\bra \hat X \ket$, with $\hat X$ the position operator.  The unitary evolution dictated by the Schr\"{o}dinger equation leaves this form unchanged, so the wave packet remains centered at zero for all times. However, if a collapse of the wavefunction occurs (triggered by a `measurement' in the standard version of Quantum Mechanics), the wave packet is displaced to one centered on a new, random value. After this reduction process, unitary evolution resumes and the wave packet begins to oscillate about zero. Repeating this process $N$ times with large $N$ leads to two contributing effects: the collapse, which tends to localize the wavefunction, and the oscillations produced by the unitary evolution. Extrapolating this behavior to the inflationary case accounts for what is found in the numerical analysis. The wavefunction associated with each mode of the field is found to be a Gaussian; while the corresponding Schr\"odinger equation is given in Eq. \eqref{schrodeq}, which corresponds to that of a harmonic oscillator with time varying frequency.

\begin{figure}[h!]
 \centering
 \includegraphics[angle=270,width=0.5\textwidth]{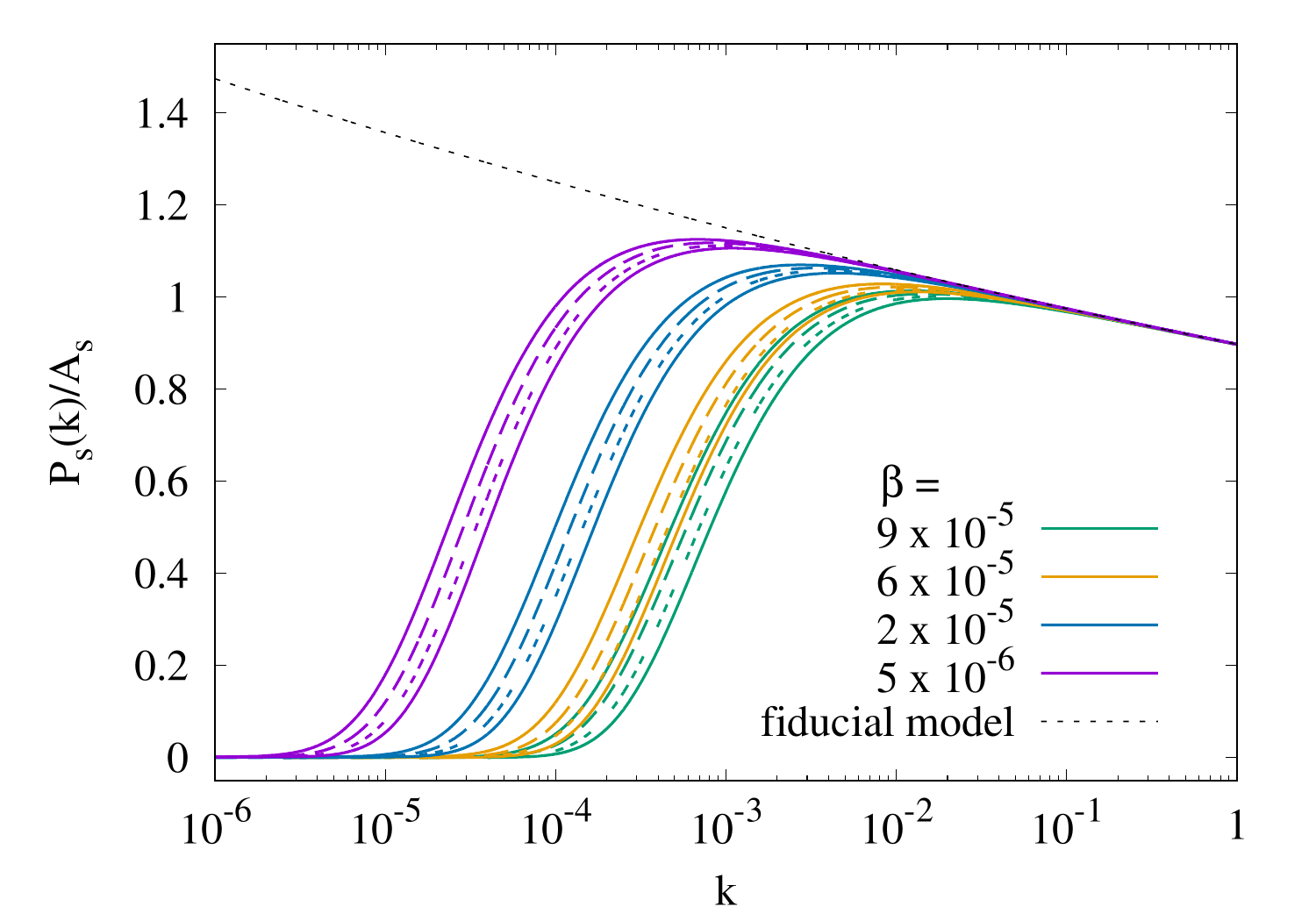}
 \caption{The units of $k$ and $\beta$ are given in Mpc$^{-1}$. When $\beta$ and $\alpha$ adopt nontrivial values, distinctive features appear in the power spectrum. Starting from the lowest $k$, which would contribute the most to generating eternal inflation, a significant departure from the fiducial model gradually arises. From left to right, $\alpha = 5$ (solid line), $6$ (long dashed), $7$ (short dashed), and $8$ (solid) for each value of $\beta$ shown in the legend. }
 \label{fig:PdeKvar_nocond}
 \end{figure}

 Moving forward in our analysis, we expect distinct features to arise when $\alpha$ and $\beta$ are nonzero. Figure \ref{fig:PdeKvar_nocond} shows the effect on the primordial power spectrum, Eq. \eqref{PSfinal}, of varying these free parameters. The changes are now apparent: a significant departure from the fiducial model at low $k$ (corresponding to larger angular scales), while for increasing $k$ there is excellent agreement. Therefore, we anticipate that these theoretical predictions will affect the lower multipoles $\ell$ of the CMB but will not appear at high $\ell$.


Furthermore, in Fig.\,\ref{fig:PdeKvar_nocond} we also observe an interesting effect of the CSL model on the primordial spectrum, namely a strong suppression at low $k$. The latter is a consequence of the parameterization \eqref{lambdaketernal}. For example, if $k \ll \beta$, then one has
\begin{equation}
	\lambda_k \simeq \lambda_0 k \gamma_k,
\end{equation}
where $\gamma_k \equiv (k/\beta)^{\alpha}$, so $\gamma_k \ll 1$. Consequently, in this case, $\gamma_k$ will suppress the amplitude of the power spectrum in Eq. \eqref{PSfinal}. Thus, we can explain the behavior of the functions in Fig. \ref{fig:PdeKvar_nocond}. In particular, if $k \ll \beta \simeq 10^{-4}\,\mathrm{Mpc}^{-1}$ e.g. in the green plots, then $\gamma_k \ll 1$ which strongly suppresses the spectrum. In the opposite regime, $k \gg \beta$ implies $\lambda_k \simeq \lambda_0 \, k$, and as a consequence, our predicted $P(k)$ approaches the fiducial one. For other values of $\beta$ shown in Fig. \ref{fig:PdeKvar_nocond}, the reasoning is the same. 

Therefore, we can interpret the previous result as primordial inhomogeneities progressively ``switching on'' and beginning to shape the primordial power spectrum until it stabilizes around the fiducial prediction. In fact, the previous behavior was expected from the proposed parameterization \eqref{lambdaketernal}. As originally argued in Ref. \cite{RL2023}, modes with $k \ll \beta$ are those we expect to play an important role in generating eternal inflation, and are those whose amplitude the present model attempts to suppress.

\begin{figure*}[!ht]
 \centering
 \includegraphics[width=0.85 \textwidth]{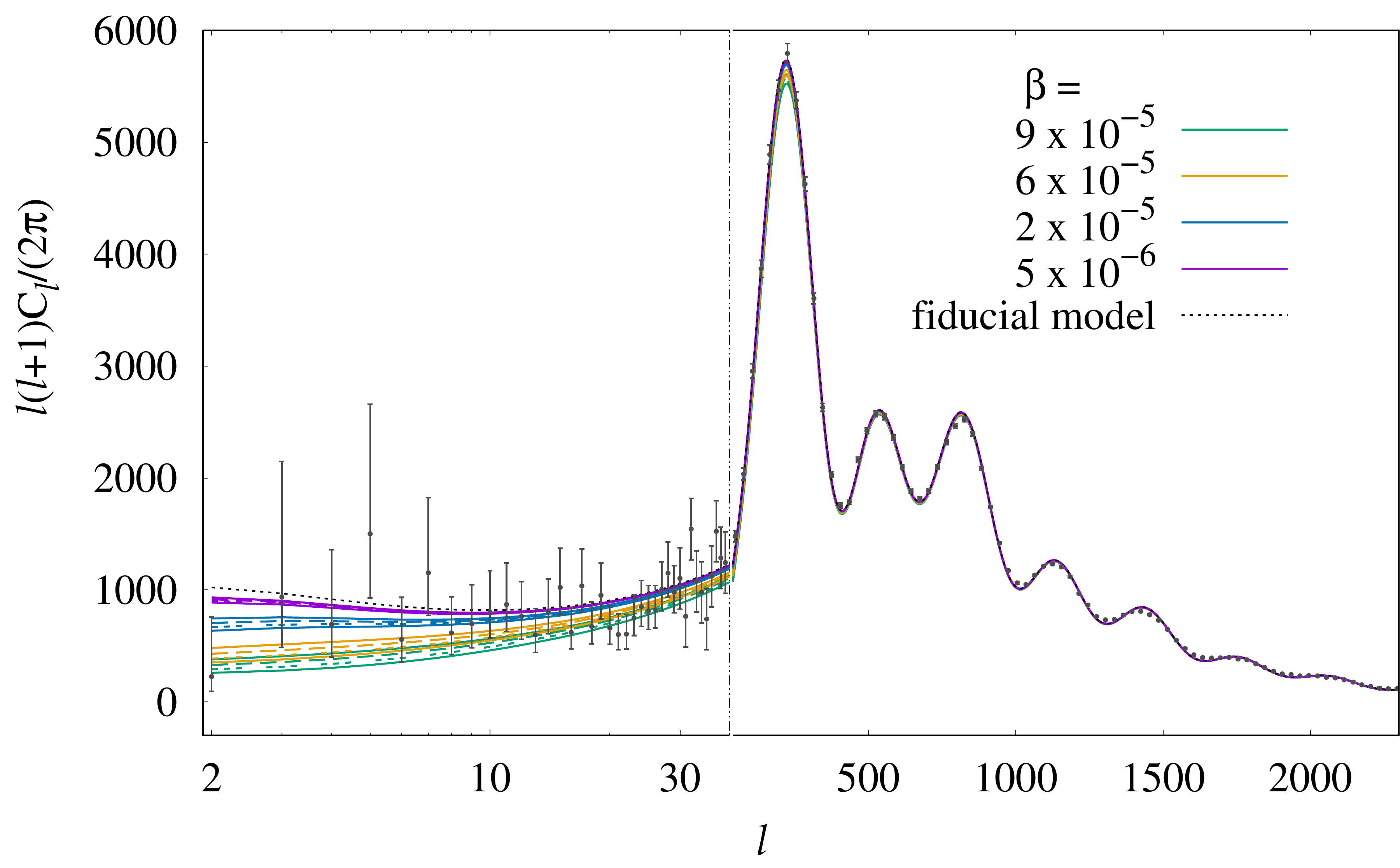}
 \caption{CMB angular anisotropies for $\alpha$ are shown for values of $5$ (solid line), $6$ (long‐dashed line), $7$ (short‐dashed line), and $8$ (solid line). A logarithmic scale was used for multipoles $2 \leq \ell \leq 40$, highlighting the main changes introduced by our model, while a linear scale was used for $41 \leq \ell \leq 2500$. It is important to note that the behavior at low $l$ is where the effect of CSL becomes evident. The parameter $\beta$ (specified in the legend) is in Mpc$^{-1}$ and takes the same values as those previously shown in the power spectrum $P(k)$. A black dashed line denotes the fiducial model, while the gray points with error bars represent the {\em Planck} data.}
 \label{fig:Cls_nocond}
 \end{figure*}

We now introduce the theoretical prediction $P(k)$ of our model in the code {\sc camb} \cite{CAMB}\footnote{Original code is available at \href{https://github.com/cmbant/CAMB}{https://github.com/cmbant/CAMB}}, which uses a line-of-sight method to compute cosmic microwave background anisotropies. The results are shown in Fig.~\ref{fig:Cls_nocond}, where the same parameter values tested for the primordial power spectra (Fig.~\ref{fig:PdeKvar_nocond}) are displayed. In this case, the {\em fiducial model}, depicted by a dashed black reference line, corresponds to a standard cosmology with the following parameters: The Hubble factor today $H_0 = 67.32$ km/s/Mpc; $\Omega_b h^2 = 0.02237$ and $\Omega_c h^2 = 0.1200$ represent baryonic and cold dark matter densities respectively, where $h=H_0/100$ [km/s/Mpc]; the reionization optical depth $\tau_{\rm reio} = 0.0544$; the spectral scalar index $n_s = 0.9649$ and the primordial scalar power spectrum amplitude $A_s = 2.1 \times 10^{-9}$.


Exactly as expected, the CSL model departs from the standard prediction at low multipoles ($l<40$), which are plotted on a logarithmic scale for better visualization. {\em Planck} data \cite{Planckcls13} are included for reference, indicating that the variations introduced by our model remain fully compatible with observational data.

This departure is interesting for future study, as it approaches the fiducial model from below and may help to resolve a long-standing issue in cosmology: the low-multipole anomaly (see, for example, \cite{2011APh....34..591S,Planck:2019evm,Copi:2010na}). This deficit of power observed at large angular scales relative to the prediction of the $\Lambda$CDM model is a persistent, unexplained feature in the observed spectrum. What is particularly puzzling is that this anomaly is found across successive all-sky experiments—COBE, WMAP, and Planck \cite{Billi_2024,stv1143,Schwarz:2015cma}. One avenue for addressing the problem involves introducing novel mechanisms operating during the inflationary epoch \cite{WMAP:2003elm}, which is precisely what our model proposes.

To further test our model, we can estimate the values of the free parameters introduced by CSL. This can be achieved through Markov chain Monte Carlo (MCMC) analysis using the latest {\em Planck} 2018 likelihood code \cite{planck2018like}, i.e. the legacy temperature and polarization likelihood {\em Planck} TT,TE,EE + lowE + lensing.\footnote{Code available at \href{https://pla.esac.esa.int}{https://pla.esac.esa.int}.} The latter follows the CMB likelihood naming convention adopted by the \textit{Planck} papers \cite{planck2018like}: \textit{Planck TT} labels the likelihood formed using only the temperature data, spanning the multipole range $2 \leq \ell \leq 2500$; \textit{Planck TE} and \textit{Planck EE} labels the likelihood formed using exclusively the \textit{TE} power spectrum from $30 \leq \ell \leq 2000 $ and the \textit{EE} power spectrum respectively. Consequently, \textit{Planck TT,TE,EE} labels the combination of \textit{Planck TT, Planck TE} and \textit{Planck EE}, taking into account correlations between the \textit{TT, TE, EE} spectra at $\ell >29$, and de contribution of \textit{TT} for $\ell \le 29$. Additionally, \textit{lowE} labels the likelihood formed using the \textit{EE} power spectrum over $2 \leq \ell \leq 29$ and \textit{lensing} labels the lensing likelihood, which includes lensing effects. We have implemented the necessary adaptations in the {\sc CosmoMC} software\footnote{Original code can be obtained at \href{https://github.com/cmbant/CosmoMC}{https://github.com/cmbant/CosmoMC}.} \cite{COSMOMC}.


As seen in Fig. \ref{fig:Cls_nocond}, varying $\alpha$ over a wide range has much less impact than a small change in $\beta$. Consequently, the results of the statistical analysis shown in Fig. \ref{figcond} (blue plots) indicate that the data do not favor a specific value of $\alpha$ but do constrain $\beta$. The priors chosen for the CSL free parameters are $\alpha = [2,8]$ and $\beta = [0,1]$. At the 68\% confidence level, we find $\alpha < 5.05$ and 
$\beta = \bigl(3.19^{+0.54}_{-3.2}\bigr)\times10^{-5}\,\mathrm{Mpc}^{-1}$. At the 99\% limit, $\beta < 1.13\times10^{-4}\,\mathrm{Mpc}^{-1}$. From the previous analysis we conclude that it is fair to say that $\beta = 10^{-3}\,\mathrm{Mpc}^{-1}$ remains an excellent upper bound according to the data.

Up to this point, we have tested a feasible interval of possible values and let the data freely choose the preferred value within that range. In other words, we have not yet imposed any physical conditions on $\alpha$ and $\beta$. However, if we wish to solve the eternal inflation problem, not every combination of $\alpha$ and $\beta$ is admissible. The next step in the analysis is to construct and apply that restriction.

\section{ Avoiding Eternal Inflation: Observational Constraints}\label{Sec4}

As we have discussed at the beginning of Section \ref{Sec:EternalCSL}, eternal inflation can occur when $\Delta_{\text{sto}}\phi_{0}>\Delta_{\text{class}}\phi_{0}$. Thus, if the CSL model is to address this problem, it must reverse that inequality. In Ref.~\cite{RL2023}, an estimate was calculated for the classical displacement $\Delta_{\text{class}}\phi_{0} \simeq 2 m^2 \phi_0 e/(3 H^2)$, as well as for the stochastic fluctuation,

\begin{equation}
\begin{split}
\Delta_{sto}\phi_{0}&=\sqrt{\overline{\bra\delta \phi\ket^2}}\\
&a(\eta)^{-1}\bigg[\int_{0}^{k_{c}}dk 4\pi k^2 \\
&\times \bigg( -\frac{\lambda_{k}\tau^{3}(\tau+\eta)}{\eta^3 k^2}\bigg)^{1/2} \bigg]
\end{split}
\end{equation}

where $k_c$ is an appropriate cut-off chosen as $k_c = 2 \pi /\tau$ and $\lambda_k$ is given in Eq. \eqref{lambdaketernal}. 

The analysis of Ref. \cite{RL2023} indicates that the eternal inflation problem can be avoided, i.e $\Delta_{\text{sto}}\phi_{0} < \Delta_{\text{class}}\phi_{0}$, if the following condition holds:
\begin{equation} \label{condifin}
 10^{15}\frac{1}{\alpha +2}\left( \frac{2\pi}{\beta \tau} \right)^{\alpha}<1.
\end{equation}
The derivation of expression \eqref{condifin} takes into account the same numerical values discussed in Section \ref{Sec:CSLobs}. In particular, $|\tau| = 10^8$ Mpc$^{-1}$, $\lambda_0|\tau|=10^3$; the typical energy scale during inflation $V^{1/4}=m^{1/2}\phi^{1/2}\simeq10^{-4}\text{M}_P$ (so $H\simeq m\phi/\text{M}_P $ ); and that inflation lasts $65$ e-foldings. Another important point to mention is that the original derivation of \eqref{condifin} in Ref. \cite{RL2023} assumed an upper bound $\beta \leq 10^{-3}$ Mpc $^{-1}$ based on an ``informed guess.'' However, the analysis of Section \ref{Sec:CSLobs} shows that the bound $\beta\leq 10^{-3}$ Mpc $^{-1}$ is supported by observational evidence.

\begin{figure}[t!]
 \centering
 \includegraphics[width=0.5\textwidth]{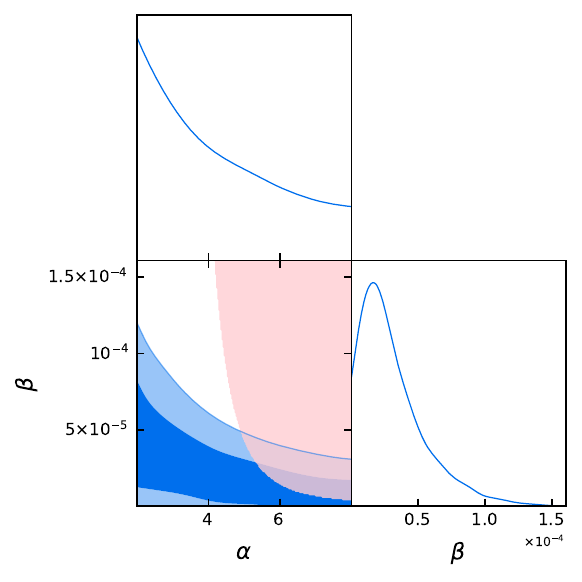}
 \caption{In blue we depict the posterior probability distributions of $\alpha$ and $\beta$, together with their joint probability confidence levels at $68\%$ and $95\%$. The light red shaded region indicates where the condition for avoiding the eternal inflation problem is satisfied. It can be seen that setting $10^{-3}$ Mpc$^{-1}$ as an upper bound for $\beta$ is both a valid and conservative choice.}
 
 \label{figcond}
\end{figure}

Our next task is to explore whether the condition of Eq. \eqref{condifin} is compatible with observational data. To address this question, Fig. \ref{figcond} superposes the region determined by the condition onto the results obtained in the previous Section, i.e. without restrictions. There is a wide area that shows incompatibility, but for lower values of $\beta$ there is overlap for both the $68\%$ and $95\%$ confidence levels.

From the discussion above, we now have certainty that CMB data validate a region in the model's parameter space. We then proceed to impose the physical condition \eqref{condifin} on the values that $\alpha$ and $\beta$ are allowed to take in the following statistical analysis.

\begin{figure}[h]
 \centering
 \includegraphics[width=0.5\textwidth]{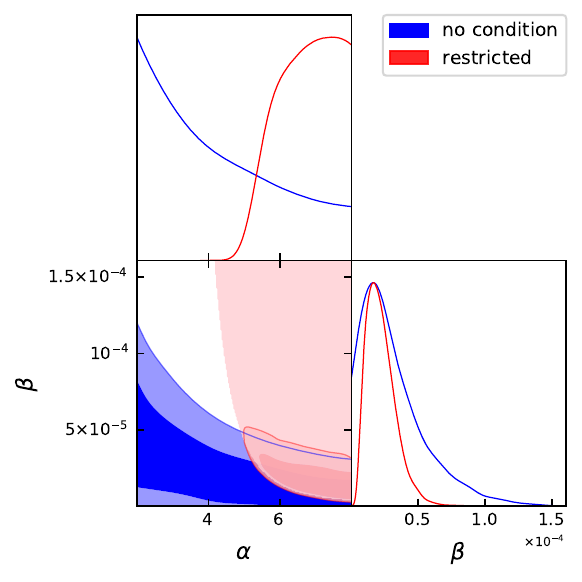}
 \caption{Imposing the theoretical condition to avoid the eternal inflation problem and taking $\beta \le 10^{-3}$ Mpc$^{-1}$ leads to much more stringent constraints. The estimates are $\beta = 2.2 \times 10^{-5}$ Mpc$^{-1}$ and $\alpha > 6.28$ ($68
 \%$ confidence level).
 The plot shows the $68\%$ and $95\%$ confidence level contours for the posterior probability of $\alpha$ and $\beta$: in blue, the results where no {\em a priori} condition was imposed; in red, the results when requiring the criterion for avoiding the eternal inflation problem to be satisfied. The light red shaded area denotes the region where this condition is met.} 
 \label{fig:sombreadoconanalisis}
\end{figure}

 Imposing the no eternal inflation condition \eqref{condifin} on the allowed values of our parameters leads to much more stringent constraints for $\alpha$ and $\beta$. The statistical analysis was re-run with the same data as before: {\em Planck}'s legacy temperature and polarization likelihood \cite{planck2018like} (\textit{Planck} TT, TE, EE + lowE + lensing). Priors were kept the same as before. However, we ensured that the code {\sc CosmoMC} was not allowed to proceed further with the algorithm, if the condition in Eq. \eqref{condifin} was not satisfied. The results comparing the restricted and unrestricted analyses for $\alpha$ and $\beta$ are shown in Fig. \ref{fig:sombreadoconanalisis}.

\begin{figure*}[!ht]
	\centering
	\includegraphics[width=0.8 \textwidth]{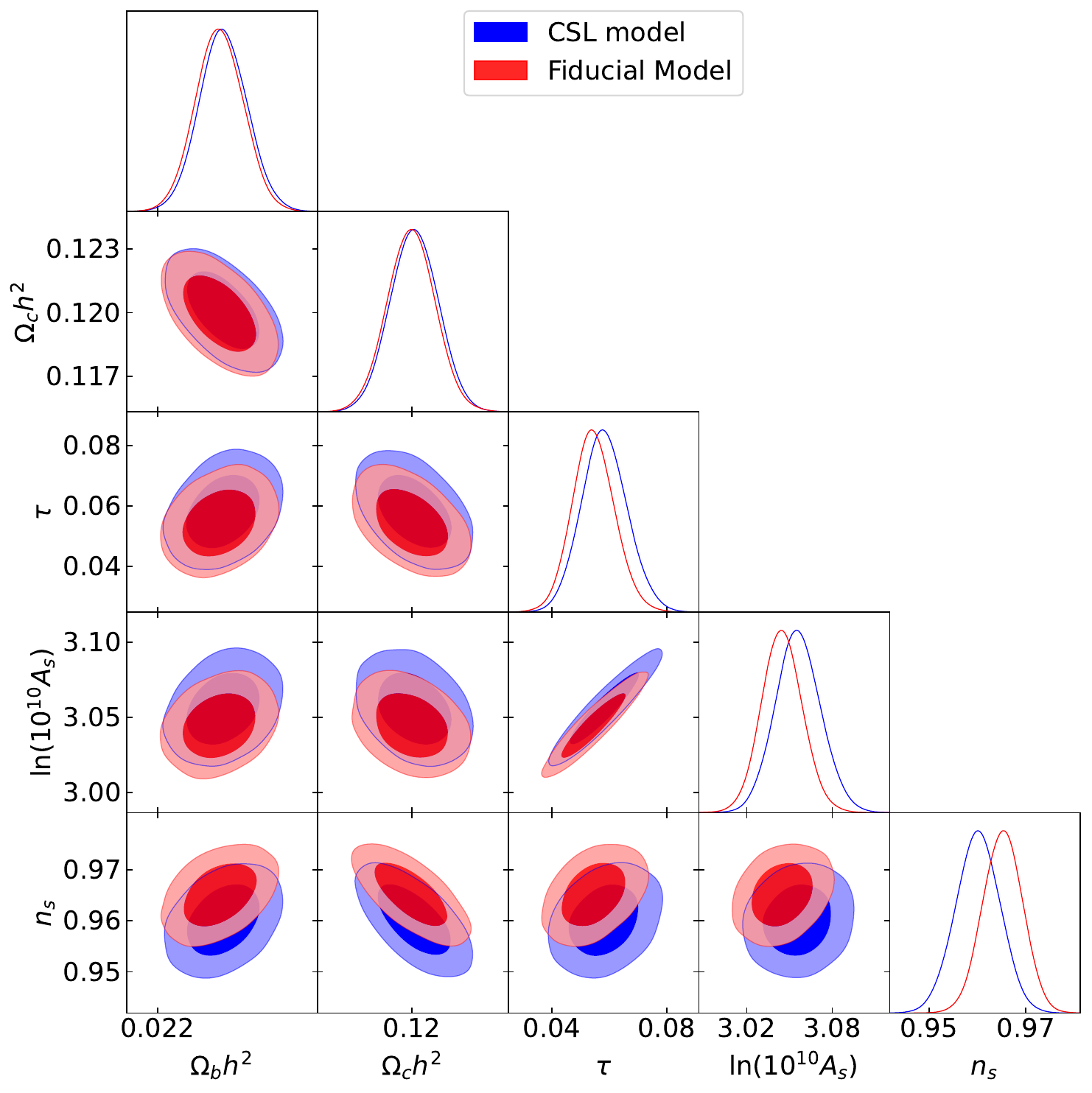}
	\caption{Posterior probability of the main cosmological parameters estimated with our model (blue) in comparison with the $\Lambda$CDM model (red). Although there is a slight shift in some of them, it is not statistically significant.}
		\label{resultados}
\end{figure*}

In order to complete our analysis, we present the results for the remaining cosmological parameters using the CSL model together with the condition \eqref{condifin}. Additionally, we again employed the fiducial model as a reference, which is simply the $\Lambda$CDM model without modifications. Figure \ref{resultados} shows the posterior probability distributions for the remaining cosmological parameters in both models, considering the $68\%$ and $95\%$ joint probability contours. Parameters such as the present day baryon and cold dark matter densities ($\Omega_b h^2$ and $\Omega_c h^2$) show no variation with respect to their estimates within the fiducial model. The reionization optical depth ($\tau_{\rm reio}$), the scalar spectral amplitude ($\ln (10^{10}A_s)$), and the scalar spectral index ($n_s$) exhibit slight shifts with no statistical significance. Since the main impact of implementing our CSL model is encoded through the primordial power spectrum, it is expected that $A_s$ and $n_s$ are affected. Due to its correlation with the scalar spectral amplitude, $\tau_{\rm reio}$ also presents a slight but not significant shift. For a quantitative comparison of the previous discussion, Table \ref{tab:resultados} presents the numerical estimates from the statistical analysis, considering a $68\%$ confidence level.

\begingroup
\renewcommand{\arraystretch}{1.10} 
\begin{table}[h]
  \centering
  \begin{tabular}[h!]{ l c c  }
    
 Parameter &  CSL & fiducial\\
\hline
{\boldmath$\Omega_b h^2   $} & $0.02240\pm 0.00015 $ & $0.02238\pm 0.00015  $\\

{\boldmath$\Omega_c h^2   $} &  $0.1201\pm 0.0012$ & $0.1200\pm 0.0012    $\\

{\boldmath$\tau_{\rm reio}           $} &  $0.0583\pm 0.0081   $ & $0.0545\pm 0.0075   $\\

{\boldmath${\rm{ln}}(10^{10} A_s)$} &  $3.056\pm 0.016  $ &  $3.045\pm 0.014   $\\

{\boldmath$n_s            $} & $0.9601\pm 0.0046 $ & $0.9652\pm 0.0041$\\

$H_0                       $ & $67.34\pm 0.54      $  & $67.37\pm 0.54$\\ 

{\boldmath$\alpha         $} & $>    6.28            $ &  $--$\\

{\boldmath$\beta              $} & $\left(\,2.20^{+0.63}_{-1.3}\,\right)\cdot 10^{-5}$ &  $--$ \\
\hline
  \end{tabular}
\caption{Analysis constraints for the models parameters using the data set \textit{Planck (2018)} + lensing. Quoted intervals correspond to $68\%$ confidence level intervals}
  \label{tab:resultados}
\end{table}
\endgroup

Summarizing the analysis of the present section, we find that the statistical analysis yields  $\beta = \left(\,2.20^{+0.63}_{-1.30}\,\right)\times 10^{-5}$ Mpc$^{-1}$ and $\alpha > 6.28$ at the $68\%$ confidence level. This confirms the hypothesis that our model is not only successful in avoiding eternal inflation but is also compatible with the latest CMB observational data.

Regarding $\alpha$, only a lower bound could be set because the collapse rate is not very sensitive to changes in this variable. In the case of $\beta$, we were able to obtain a numerical estimate. 
Analyzing the posterior probability of this parameter, we think it is possible to achieve better constraints. The errors obtained in the estimation may be attributable to the manner in which the parameter space is explored. Future work could focus on identifying a more efficient approach to navigating the parameter space, in order to improve this estimation using different numerical techniques.

\section{Summary and Conclusions}\label{Sec5}

The incorporation of the inflationary ``paradigm'' into the standard cosmological model has undoubtedly encountered great successes; however, these come together with various problematic aspects. On the one hand, we have the status of the primordial gravitational waves,  for which the constraints of empirical results and standard predictions have led to the dismissal of a large set of models, and the issue of eternal inflation, which is often left as an unresolved task for future treatment. On the other hand, there are the lingering conceptual questions (for which no general consensus seems to have emerged) regarding the lack of clarity or even inability of the standard framework to account for the emergence of the primordial inhomogeneities and anisotropies in a context where that symmetry is present in the initial state and the dynamics has no element to break it. 
This work, as well as several previous ones \cite{Perez:2005gh, Canate:2012nwv, piarunning, Martin:2012pea, Leon2021x, commentshort}, emphasizes that it is necessary to modify the dynamics evolution in order to account for the breaking of symmetry in the Bunch-Davies vacuum and then explain the emergence of the primordial cosmic structure. Interestingly, the treatments based on semi-classical gravity together with spontaneous collapse theories offer both a clear resolution of the problem and a natural explanation for the observational absence of primordial gravity waves\cite{nobmodesbig}. A central aspect of those discussions is the inappropriateness of identifying quantum uncertainties (or two-point functions) as measures of stochastic (temporal-space dependent) fluctuations. That realization indicates, at the same time, the need to reevaluate the eternal inflationary scenarios.

However, as it was pointed out in \cite{RL2023}, while the reassessment of the nature of the quantum uncertainties seems to remove the eternal inflation issue, the incorporation into the cosmological model of spontaneous collapse dynamics could make the issue reemerge. 
That is, the same kind of stochastic fluctuations in the dynamics that can account for the breaking of symmetry in the Bunch-Davies vacuum could, in principle, lead to an eternal inflation scenario. Nevertheless, as shown in the present work, the detailed characteristics in the dynamic evolution equations offer a path to avoid \textit{eternal inflation}, and offer a satisfactory account for the emergence of the seeds of cosmic structure with an appropriate primordial spectrum.

The manner of working with these theories in cosmology has been the following: first, clarifying the conceptual aspects. Then, attempting to make the theoretical model more robust, and finally, ensuring the viability of the model by confronting it with observational data. In some cases, the exploration involves examining whether certain predictions under the new approach for the cosmological situation would differ substantially from the standard ones. This is the case, for instance, concerning the prediction of primordial gravitational waves \footnote{The predictions in the CSL model combined with semiclassical gravity change for the B-modes, showing a suppression \cite{Palermo:2022dim} \cite{nobmodesbig}.}.
Following that spirit, the primary objective of this manuscript was to determine whether the theoretical model proposed in Ref. \cite{RL2023}, which studies the eternal inflation problem in light of the CSL model, is compatible with the CMB data.


The exploration of the problem began with a search for a viable interval for the parameters' values,   following with a study to select the values preferred by the data, without imposing any further restrictions “by hand” on the two new parameters, $\alpha$ and $\beta$, which characterize the collapse rate $\lambda_k$ of the CSL inflationary model [see, Eq. \eqref{lambdaketernal}] \footnote{As is stated in \cite{RL2023}, this parameterization has the feature that accounts for the structure (modes with small wavelength) and also for the eternal inflation (linked with $\lambda_{phys}$ large)}.

In the first part of our analysis, no restrictions were imposed on $\alpha$ and $\beta$; the aim was to explore only the parametrization of the collapse rate through these two parameters, without considering the eternal inflation problem, and simply to determine the values allowed by the data. The analysis was carried out using the \textit{Planck (2018)} legacy temperature and polarization data, together with the lensing likelihood. In this unrestricted part, we considered two scenarios. The first corresponds to $\beta=0$, which is the same case studied in \cite{piarunning}, and which reproduces the usual power-law shape at leading order. Nevertheless, small-scale oscillatory features, a characteristic signature of the CSL collapse mechanism, persist. However, these oscillations have amplitudes low enough that they remain unobservable in the angular power spectrum. In the second case, we considered nonzero values of both $\alpha$ and $\beta$. The changes become evident at larger angular scales, while at small scales (higher $k$ values) the prediction remains consistent with the fiducial model. Altogether, this behavior constitutes a promising scenario in addressing the low multipole anomaly \cite{2011APh....34..591S}. 

The results of the statistical analysis suggest that the data do not favor a specific value of $\alpha$, but they do constrain $\beta$, for which we estimated $\beta \leq 10^{-3}$ Mpc$^{-1}$. Finally, the solution region plotted in \cite{RL2023} was compared with the unrestricted region, and we found that there is an overlap for lower values of $\beta$, within the $95\%$ confidence level.


Then, in the next part of the analysis, we imposed the condition given in Eq. \eqref{condifin}, under which the eternal inflation problem is avoided. This condition restricted the $(\alpha,\beta)$ parameter space, defining a validity region for these parameters, as shown in Fig. \ref{figcond}. After performing the statistical analysis with this restriction on the allowed parameter values, we obtained the estimates $\alpha \gtrsim 6$ and $\beta \sim 10^{-5}$ Mpc$^{-1}$. The overall effect on the remaining cosmological parameters is not statistically significant.

From the discussion above, we conclude that our proposal is not only compatible with the latest observational data, but also capable of avoiding the eternal inflation scenario. The proposed model requires only two additional parameters, which can be successfully estimated using the latest Planck legacy data. It is worthwhile emphasizing that our approach goes well beyond the realm of theoretical or philosophical inquiry and yields concrete observational predictions which can be contrasted with empirical input leading to a satisfactory picture both at the conceptual level and and the level of observational adequacy.

\appendix
\section{CSL in a nutshell}
\label{apendiceRA}

{Collapse theories have been designed to address the measurement problem in Quantum Mechanics. We remind the reader that this problem lies in the fact that quantum theory, as presented in textbooks, invokes two different evolution rules \cite{RPenroseUyR}, one completely deterministic, embodied by Schr\"odinger’s equation, and another one of stochastic nature, that is supposed to apply when a measurement is performed on the quantum system. The issue is that the theory is, in general, quite obscure regarding the general rules determining when one or the other evolution laws apply, that is, what exactly counts as a measurement\footnote{{Simple proposals such as ``all interactions are measurements'' are ruled out by interference experiments.}}. The difficulty becomes apparent when considering Bell's rhetorical question \cite{bell1}: \enquote{{What exactly qualifies some physical systems to play the role of ``measurer''? Was the wavefunction of the world waiting to jump for thousands of millions of years until a single-celled living creature appeared?. . .}} In fact, this could be most alarming, adding questions like:  when does a particular measurement take place?, what quantity is being measured?, etc. }

{Collapse theories work by unifying the two evolution rules and removing any {\it ``apriori''} distinction in the manner of treating the system of interest and the apparatus; all the physical systems are treated in quantum mechanical terms.  The  ``special'' role of apparatus emerges (rather than being postulated) from the universal evolution law as a result of the large number of degrees of freedom in it, and the entanglement between these and those of the system of interest caused by their interactions. The basic idea is that the effect of the modified general dynamics, involving departure from unitary, differs in a negligible way when a very small number of elementary degrees of freedom are involved, but might become dominant when a vast number of degrees of freedom are involved. {This process is called {the amplification mechanism}, and is a generic feature of the collapse models. }} 

{Collapse theories have been originally constructed for applications in the non-relativistic many-particle quantum mechanics. The first realistic version of these theories is the GRW  proposal \cite{GRW}, which involves discrete spontaneous reduction of the wave function for the particle's position degrees of freedom. A very attractive alternative is the Continuous Spontaneous Localization \cite{pearlemisc},  which naturally allows the treatment of identical particles, and involves a Wiener-like continuous reduction process.}

{It is an important remark that, in the literature, there exist different versions of the same CSL theory; many of them (the non-relativistic cases) are described in \cite{pearlemisc}. As one might expect, the general idea is the same, but some of the main differences concern the choice of the collapse operator and the collapse parameter.
The ``most general'' version of the theory (in the sense that it allows for a generic form of the collapse operator and its associated collapse rate) is presented in \cite{pearlemisc}.}

{The mathematical framework consists of two equations: \textrm{i)} a stochastic differential evolution equation, whose solution takes the form: 
  \begin{equation}\label{CSL1}
{{ |\psi,t\rangle_w=\hat {\cal T}e^{-\int_{0}^{t}dt'\big[i\hat H+\frac{1}{4\lambda}[w(t')-2\lambda\hat A]^{2}\big]}|\psi,0\rangle.}}
\end{equation}
}
{where $\hat {\cal T}$ s the time-ordering operator, $A$ the collapse operator, $w(t)$ the random classical function of time.}

{The second equation  is the probability rule for the realization $\tilde{w}(t)$ of  an element of  stochastic class $\lbrace w(t) \rbrace$, restricted to lie  in the ``tube''  in the plane $w-t$ defined by the  tolerance intervals $\text{d}w (t_i)$ around the values $\tilde{w}(t_i)$.}

\begin{equation}\label{CSL2Prob-Rule}
{PDw(t)\equiv{}_{\tilde w} \langle\psi,t|\psi,t\rangle_{\tilde w} \prod_{t_{i}=0}^{t}\frac{dw(t_{i})}{\sqrt{ 2\pi\lambda/dt}},}
\end{equation} 

{A particular type of these models was written in the context of a single particle in non-relativistic quantum mechanics, in this setup the collapse operator is a smeared version of the position operator, \footnote{{This operator can be characterized formally in terms of the particle position basis ${|x\rangle}$ writing ${\hat{X_{r}} \equiv  N \int dx|x+y\rangle x\langle x-y | e^{-\frac{||x-y||^{2}}{2r^2}}}$.}} { with smearing scale $r_c$.  When  $r_c \to 0$ this operator becomes the usual position operator, i.e. $\hat{A}=\hat{X_r}$ Therefore, the CSL model is characterized by two parameters: the collapse rate $\lambda$ (typically very small, $\sim 10^{-17} \mathrm{s}^{-1}$, which guarantees that collapses are extremely rare for microscopic systems) and the localization length $r_c$ (also small, $\sim 10^{-7} \mathrm{m}$, which ensures an efficient localization for macroscopic systems).
In addition, the CSL model has been subjected to experimental scrutiny, since its predictions differ from those of standard quantum mechanics for certain physical systems \cite{BassiNat}. The resulting empirical bounds have been derived from a variety of observations and experiments, including astrophysical bounds \cite{Ocampo2024,LochanDas2012}, spontaneous X-ray emission \cite{sandro2017,Majorana2022}, matter--wave interferometry \cite{Gasbarri2021}, and gravitational wave detectors \cite{carlesso2016, lisa2017}. } }

{ The extension of collapse models into the context of relativistic quantum field theories is far from trivial, in part,   because in such setting there is no fully satisfactory candidate for a position operator \cite{pearlemisc} \footnote{{There are various other issues that arise such as the need to rely on relativistic versions of a smearing procedure (\cite{Myrvold:2021jse} for a discussion of those issues, and \cite{Bedingham:2019inf} for a proposal)}.}. {The problem is aggravated further when considering quantum field theory in curved spacetimes because particles are not fundamental objects (only quantum fields are \cite{Wald94}). Consequently, the notion of localization differs substantially from the particle-based picture: while particle localization is associated with position and momentum uncertainties, hence introducing naturally a length scale $r_c$, field localization is instead related to uncertainties in the field values and their conjugate momenta, suggesting a parameter with different physical characteristics  (e.g. physical dimensions). Therefore, there is, at this time,  no clear and straightforward connection between the two notions of state reduction required for these different contexts. Others, making use of different choices, obtain different results \cite{jmartin, Das2013}. The issue has been discussed in \cite{Bengochea:2023dep}. A similar issue arises for the collapse rate parameter $\lambda$, which is constrained by laboratory experiments but must be extrapolated to the early universe, a regime involving multiple phase transitions and poorly understood epochs. For these reasons, there is no compelling basis for a simple correspondence between CSL parameters defined at laboratory scales and those relevant in cosmological settings.} }

{ {Nevertheless discussion above does not imply that a consistent phenomenological description in cosmology is unattainable.} The cosmological setting is one in which, in practice, several of these conceptual difficulties can be bypassed, owing to the high degree of symmetry (even if only approximate) of the background spacetime and to the existence of preferred time parameters naturally associated with cosmic evolution. However, even then, multiplicity of options for the collapse operators $A$ exists (see, for instance, the discussion in \cite{Leon2021x}  and the exploration of the results of some specific choices in \cite{Martin:2021lje,Bengochea:2020qsd, TPSingh}).}
{The version of the CSL theory used in this manuscript is described by equations \ref{CSL1} and \ref{CSL2Prob-Rule}, using the field\footnote{This proposal corresponds to taking the scalar field at each point as the collapse operator, and therefore replacing the sum over degrees of freedom $i$ by the spatial integral $\boldsymbol{\int \sqrt{h} d^{3}x}$.}  and its conjugate momentum as collapse operators $\hat{A}$. This particular formulation of the theory was introduced in \cite{Canate:2012nwv} to account for the primordial power spectrum. That work considers various choices for the collapse operator $\hat{A}$ and found that the form of $\lambda$ that ensures an essencially scale invariant spectrum is required to become a function \footnote{{In order to preserve scale invariance in the power spectrum, a specific functional form for the collapse rate $\lambda$ is chosen.}} of the field modes $k$. In particular, when the collapse operator is taken to be the rescaled field $a\delta\hat{\phi}$, the form that is needed is $\lambda(k)=\lambda_{0}k$. For typical values of the inflationary model parameters the value for the parameter $\lambda_{0}$ turns out to be $\lambda_{0}=10^{-19}\text{s}^{-1}$ a value which is interestingly close to the range of values contemplated for the parameters appearing in non-relativistic quantum mechanical versions of the theory \cite{XENON2025}. On the other hand, in \cite{RL2023}, a crucial modification of the parametrization $\lambda(k)=\lambda_{0}k$ was introduced  (see \eqref{lambdaketernal}; however, the value; however the value $\lambda_{0}=10^{-19}\text{s}^{-1}$ was retained) in order to address simultaneously the emergence of primordial structure and the problem of eternal inflation.} 

\section{Solving the CSL equations in the primordial power spectrum}
\label{apendiceA}
Our expression for the primordial power spectrum is given in Eq. \eqref{masterPS2}, which we reproduce again here:
\begin{equation}\label{masterPS2app}
	\mP_{s} (k) = \frac{k^3}{4 \pi^2 M_P^2 a^2 \epsilon_1^2} \left( \overline{\bra \hat y_{\nk}^2 \ket} - \frac{1}{4 \textrm{Re} [A_k(\eta)]} \right).
\end{equation}

We focus on the first term in \eqref{masterPS2app}, namely $\overline{\bra \hat y_{\nk}^2 \ket}$, and introduce the following definitions:
\begin{equation}\label{defQRS}
	Q \equiv \overline{\bra \hat y_{\nk}^2 \ket}, \quad R \equiv \overline{\bra \hat p_{\nk}^2 \ket}, \quad S \equiv \overline{\bra \hat p_{\nk} \hat y_{\nk} + \hat y_{\nk} \hat p_{\nk} \ket}.
\end{equation}

The evolution equations for $Q$, $R$, and $S$ can be derived from the CSL dynamics. Specifically, for any operator $\hat O$, the following equation holds:
\begin{equation}\label{vesperadoO}
	\frac{d}{d \eta} \overline{\bra \hat O_\nk \ket} = -i \overline{\bra [ \hat O_\nk, \hat H_\nk ] \ket} - \frac{\lambda_k}{2} \overline{\bra [ \hat y_\nk , [ \hat y_\nk, \hat O_\nk ]] \ket},
\end{equation}
which governs the evolution of the ensemble average of the expectation value of $\hat O$. Applying this to our case, we obtain the following system of linear differential equations:

\begin{equation} \label{QRS}
\begin{split}
    	Q' = S - \frac{2}{\eta} (1 + \epsilon_1 + \epsilon_2/2) Q, \\
        R' = -k^2 S + \frac{2}{\eta} (1 + \epsilon_1 + \epsilon_2/2) R + \lambda_k, \\ S' = 2R - 2k^2 Q.
\end{split}
\end{equation}

The general solution consists of a particular solution to the inhomogeneous system and the general solution of the homogeneous part (where $\lambda_k = 0$). The specific solution of interest is:
\begin{equation}\label{Q}
\begin{split}
    Q(\eta) =& (-k\eta) \bigg[ C_1 J_m^2(-k\eta) + C_2 J_{-m}^2(-k\eta) +\\
    & C_3 J_m(-k\eta) J_{-m}(-k\eta) \bigg] + \frac{\lambda_k \eta}{2k^2},
\end{split}
\end{equation}
with $m = 3/2 + \epsilon_1 + \epsilon_2/2$. The constants $C_1$, $C_2$, and $C_3$ are determined by the initial conditions associated with the Bunch-Davies vacuum: $Q(\tau) = 1/(2k)$, $R(\tau) = k/2$, and $S(\tau) = 0$. Equation \eqref{Q} is exact; in the super-Hubble regime, where $-k\eta \ll 1$, it admits the approximate form:
\begin{equation}\label{Qaprox}
\begin{split}
    	Q(\eta) &\simeq \frac{\pi}{2k^2 \sin^2(m\pi)}\\
        &\times \bigg[ \frac{k}{2} - \frac{\lambda_k \tau}{2} + \frac{m \lambda_k}{k} \sin\Delta \cos\Delta \bigg]\\
        &\times \frac{2^{2m}}{\Gamma^2(1-m)} (-k\eta)^{-2m+1},
\end{split}
\end{equation}
where
\begin{equation}
	\Delta = -k\tau - \frac{m\pi}{2} - \frac{\pi}{4}.
\end{equation}

On the other hand, the second term on the right hand side of \eqref{masterPS2app}, $(4 \textrm{Re} [A_k(\eta)])^{-1}$, corresponds to the variance of the field variable, which is directly related to the width of the wave functional in \eqref{psionday}. To derive the evolution equation for $A_k(\eta)$, we differentiate \eqref{CSLevolution} with respect to time and apply the resulting operator to the wave functional \eqref{psionday}. By organizing the resulting terms according to their powers in $y$—specifically $y^2$, $y^1$, and $y^0$—the corresponding evolution equations decouple. In particular, the equation associated with the $y^2$ term involves only $A_k(\eta)$, allowing it to evolve independently from $B_k(\eta)$ and $C_k(\eta)$. Thus, the evolution equation of $A_k$ is :
 \begin{equation}\label{evolAk} 
 A_k' = - 2i A_k^2 -2 \frac{z'}{z} A_k + \frac{i k^2}{2} + \lambda_k. 
\end{equation} 
Introducing the change of variable $A_k \equiv f'/(2i f)$, Eq. \eqref{evolAk} can be rewritten as: 
\begin{equation}\label{evolfk} 
f'' + \frac{2 z'}{z} f' + q^2 f = 0
\end{equation} 
where 
\begin{equation}\label{defq}
 q^2 \equiv k^2 \left(1 - 2i \frac{ \lambda_k}{k^2}\right) = k^2 \zeta_k^2 e^{i 2 \theta_k },
\end{equation} 
and in the last term, we have used the definitions in \eqref{defzetatheta}. Also, recalling the definition of $z \equiv a \sqrt{\epsilon_1}$ and the definition of $\epsilon_2$, we have the exact relation
\begin{equation}
	\frac{z'}{z} = \mH \left(1+ \frac{\epsilon_2}{2} \right),
\end{equation}
moreover, expanding $\mH$ at the lowest order in $\epsilon_i$, we have $\mH = -(1+\epsilon_1)/\eta + \mathcal{O}(\epsilon_i^2)$, so
\begin{equation}\label{terminoz}
	\frac{z'}{z} = - \frac{1}{\eta} \left( 1 + \epsilon_1 + \frac{\epsilon_2}{2} \right) + \mathcal{O}(\epsilon_i^2)
\end{equation}
Substituting Eq. \eqref{terminoz} in Eq. \eqref{evolfk}, the equation for $f$, at the lowest order in the HFF, is
\begin{equation}
f'' - \frac{2}{\eta} \left( 1 + \epsilon_1 + \frac{\epsilon_2}{2} \right) f' + q^2 f = 0
\end{equation}
At this point, we can treat $\epsilon_1$ and $\epsilon_2$ as approximately constants. Consequently, the previous equation is a Bessel differential equation, so the solutions are linear combinations of Bessel functions. Returning to the original variable $A_k$, the solution is 
\begin{equation}\label{Aexacta}
	A_k(\eta) = \frac{q}{2i} \left[\frac{-J_{m-1} (-q\eta)-e^{-i\pi m} J_{1-m} (-q\eta)}{J_{m} (-q\eta) -e^{-i\pi m} J_{-m} (-q\eta)} \right]
\end{equation}
where $m = 3/2 + \epsilon_1 + \epsilon_2/2$ (the same as before). The solution \eqref{Aexacta} satisfies the Bunch-Davies initial condition, corresponding to $A_k(\tau) = k/2$

We now expand $A_k(\eta)$ in the limit where the proper wavelength associated to the modes becomes larger than the Hubble radius, that is, in the limit $-q\eta \to 0$ (provided that $\lambda_k \ll k^2$). However, we are only interested in the expansion on the real part of $A_k (\eta)$, this is

\begin{equation}\label{reacaso1}
\begin{split}
 &  \text{Re} \: A_k(\eta) \simeq \frac{\lambda_k}{2k(m-1)}(-k\eta) \\
   &+ \zeta_k^{2m} \frac{\sin(\pi m+ 2m\theta_k) }{\sin (\pi m)} \frac{k \pi}{2^{2m} \Gamma(m)^2} (-k\eta)^{2m-1}, 
\end{split}
\end{equation}

Using the results of Eqs. \eqref{Qaprox} and \eqref{reacaso1}, we can finally compute the primordial PS as given in \eqref{masterPS2app}, this is,

\begin{equation}\label{masterPS3app}
    \begin{split}
       & \mP (k) = \frac{k^3}{4 \pi^2 M_P^2 a^2 \epsilon_1^2} \left( Q(\eta) - \frac{1}{4 \textrm{Re} A_k (\eta)} \right) \\
&= \frac{k^3}{4 \pi^2 M_P^2 a^2 \epsilon_1^2} \bigg\{ \frac{\pi}{2k^2 \sin^2(m\pi)}\\
&\times \bigg( \frac{k}{2} - \frac{\lambda_k
	\tau}{2} + \frac{m \lambda_k}{k} \sin\Delta \cos\Delta \bigg)
 \\
&\times \frac{2^{2m}}{\Gamma^2(1-m)} (-k\eta)^{-2m+1} - \frac{1}{4} \bigg[ \frac{\lambda_k}{2k(m-1)}(-k\eta) \\
&+\zeta_k^{2m}
\frac{\sin(\pi m+ 2m\theta_k) }{\sin (\pi m)} \frac{k \pi}{2^{2m} \Gamma(m)^2}
(-k\eta)^{2m-1} \bigg]^{-1} \bigg\}.
    \end{split}
\end{equation}

At the lowest order in the HFF, the previous equation can be expressed as

\begin{equation}
    \begin{split}
        	\mP(k) &\simeq \frac{H^2}{8 \pi^2 M_P^2 \epsilon_1} (-k\eta)^{-2 \epsilon_1 - \epsilon_2}\\
            &\times \bigg\{ 1 + \frac{\lambda_k |\tau|}{k} + \frac{3 \lambda_k}{2 k^2} \sin (2 k |\tau|) \nn
	&- \bigg[ \frac{2 \lambda_k}{k^2} (-k \eta)^{-1 -2 \epsilon_1 - \epsilon_2 } \\
    &+ \zeta_k^{3 + 2 \epsilon_1 + \epsilon_2} \cos[ (3 + 2 \epsilon_1 + \epsilon_2) \theta_k] \bigg]^{-1} \bigg\}.
    \end{split}
\end{equation}

Finally, it is customary to evaluate the power spectrum at the time of ``horizon crossing'' of a pivot mode $k_\star$, i.e. $-k_\star \eta_\star = 1$; moreover, employing the standard definition of the scalar spectral index 
$n_s \equiv 1 - 2\epsilon_{1 *} - \epsilon_{2 *}$, we find the primordial PS
\begin{equation}\label{masterPSfinal}
    \begin{split}
     	&\mP(k) \simeq \frac{H_\star^2}{8 \pi^2 M_P^2 \epsilon_{1 \star}} \left( \frac{k}{k_\star} \right)^{n_s-1}\\
        &\times\bigg\{ 1 + \frac{\lambda_k |\tau|}{k} + \frac{3 \lambda_k}{2 k^2} \sin (2 k |\tau|) \nn
	&- \left[ \frac{2 \lambda_k}{k^2} \left( \frac{k}{k_\star} \right)^{n_s -2 } + \zeta_k^{4-n_s} \cos[ (4-n_s) \theta_k] \right]^{-1} \bigg\}.   
    \end{split}
\end{equation}

\section*{Acknowledgments}

The authors gratefully acknowledge the computing time granted by LANCAD and SECIHTI on the supercomputer Yoltla/Miztli/Xiuhcoatl at LSVP UAM-Iztapalapa/DGTIC UNAM/CGSTIC CINVESTAV. M.P.P. and G.L. are supported by CONICET (Argentina). Also, M.P.P. and G.L. acknowledge support from the project grant Universidad Nacional de La Plata (UNLP) I+D No. SG002. 
D.S acknowledges support from the Proyect PAPIIT No IG100124, and  a sabbatical year Fellowship
 from UNAM, through a PASPA/DGAPA and the hospitality of the Departament de Física Quantica i Astrofísica Universitat de Barcelona.


\bibliography{bibliografia_eternal}
\bibliographystyle{JHEPnew}
%
%
%



\end{document}